\RequirePackage{fix-cm}
\documentclass[twocolumn,epjc3]{svjour3} 
\smartqed  
\RequirePackage{graphicx}
\journalname{Eur. Phys. J. C}
\begin{document}

\title{A hidden-charm \mbox{\boldmath{$S=-1$}} pentaquark from the decay of \mbox{\boldmath{$\Lambda_b$}} into \mbox{\boldmath{$J/\psi \, \eta \Lambda$}} states
}


\author{A. Feijoo\thanksref{e1,addr1}
        \and
        V.K. Magas\thanksref{e2,addr1} 
        \and
        A. Ramos\thanksref{e3,addr1} 
        \and
        E. Oset\thanksref{e4,addr2} 
}

\thankstext{e1}{e-mail: feijoo@icc.ub.edu}
\thankstext{e2}{e-mail: vladimir@icc.ub.edu}
\thankstext{e3}{e-mail: angels.ramos@icc.ub.edu}
\thankstext{e4}{e-mail: Eulogio.Oset@ific.uv.es}


\institute{Departament de F\'isica Quˆntica i Astrof\'isica and Institut de Ci\`encies del Cosmos (ICCUB), 
		       Universitat de Barcelona, Mart\'i i Franqu\`es 1, 08028 Barcelona, Spain \label{addr1}
           \and
           Departamento de F\'{\i}sica Te\'orica and IFIC, Centro Mixto Universidad de Valencia-CSIC Institutos de 
           Investigaci\'on de Paterna, Aptdo.22085, 46071 Valencia, Spain \label{addr2}
}

\date{Received: date / Accepted: date}

\maketitle

\begin{abstract}
The hidden charm pentaquark $P_c(4450)$ observed recently by the LHCb collaboration may be of molecular nature, as advocated by some unitary approaches that also predict pentaquark partners in the strangeness $S=-1$ sector. In this work we argue that a hidden-charm strange pentaquark could be seen from the decay of the $\Lambda_b$, just as in the case of the non-strange $P_c(4450)$, but looking into the  $J/\psi\, \eta \Lambda$  decay mode and forming the invariant mass spectrum of $J/\psi\Lambda$ pairs. In the model presented here, which assumes a standard weak decay topology and incorporates the ha\-dro\-ni\-za\-tion process and final state interaction effects, we find the $J/\psi \, \eta \Lambda$ final states to be populated with similar strength as the $J/\psi K^- p$  states employed for the observation of the non-strange pentaquark. This makes the $\Lambda_ b \to J/\psi \, \eta \Lambda$ decay to be an interesting process to observe a possible strange partner of the $P_c(4450)$ . We study the dependence of the $J/\psi \Lambda$ mass spectra on various model ingredients and on the unknown properties of the strange pentaquark. 
\end{abstract}

\section{Introduction}

The LHCb collaboration reported recently two exotic structures in the invariant $J/\psi p$ mass spectrum of
the $\Lambda_b^0\rightarrow J/\psi K^- p$ process. These pentaquark states were named  $P_c(4380)$, with a mass of $4380\pm8\pm29$ MeV and a width of $205\pm18\pm86$
MeV, and $P_c(4450)$, with a mass of 
$4449.8\pm1.7\pm2.5$ MeV and a width of $39\pm5\pm19$ MeV ~\cite{Aaij:2015tga,Aaij:2015fea}. 
Hidden charm baryon states with similar characteristics of the states reported had already been predicted, employing a molecular picture~\cite{Wu:2010jy,Wu:2010vk,Yang:2011wz,Xiao:2013yca,Karliner:2015ina} or a quark model approach~\cite{Wang:2011rga,Yuan:2012wz}. A list of early references on pentaquark states can be seen in Ref.~\cite{Stone:2015iba}. The CERN discovery
triggered a large number of theoretical works trying to give an explanation for the two reported states.
The molecular picture was invoked in ~\cite{Chen:2015loa,Roca:2015dva,He:2015cea,Meissner:2015mza},
the diquark picture in ~\cite{Lebed:2015tna,Maiani:2015vwa,Anisovich:2015cia,Ghosh:2015ksa,latest:diquark}, QCD sum rules were used in ~\cite{Chen:2015moa,Wang:2015epa},
and  the soliton model was employed in ~\cite{Scoccola:2015nia}.
It has also been argued that the observed enhancement could be due to kinematical effects or triangular singularities~\cite{Guo:2015umn,Liu:2015fea,Mikhasenko:2015vca}. Suggestions of different reactions to observe the pentaquarks have been reported \cite{Huang:2013mua,Garzon:2015zva,Wang:2015jsa,Kubarovsky:2015aaa,Karliner:2015voa}, while explicit decay modes to elucidate their structure have also been studied in \cite{Cheng:2015cca,Li:2015gta}. Further discussions on the issue and the nature of the two $P_c$ states can be seen in Refs.~\cite{Mironov:2015ica,Burns:2015dwa} and particularly in the recent detailed review of Ref.~\cite{Chen:2016qju}.

In what respects the present work, we recall that a theoretical study  
of the $\Lambda_b^0\rightarrow J/\psi K^- p$ reaction was done in \cite{rocamai}, prior to the experimental study of Ref.  \cite{Aaij:2015tga}, predicting the contribution of the tail of the  $\Lambda(1405)$ in the $K^- p$ invariant mass distribution. The analysis of \cite{Aaij:2015tga}
contains such a contribution in agreement in shape with the predictions, where the absolute normalization is unknown. Moreover, it was shown in 
\cite{Roca:2015dva} that the distributions in the pentaquark channel, i.e. in the invariant $J/\psi p$ mass spectrum of \cite{Aaij:2015tga}, could be explained via the incorporation of the hidden charm $N^*$ states predicted in \cite{Wu:2010jy,Wu:2010vk,Yang:2011wz,Xiao:2013yca}, which are molecular states mostly made from $\bar D^* \Sigma_c$ or $\bar D^* \Sigma_c^*$ components and having a small coupling to $J/\psi p$, one of their open decay channels. 
It is unlikely that there are no partners of the states found in \cite{Aaij:2015tga}, and indeed, in \cite{Wu:2010jy,Wu:2010vk} states of spin-parity $3/2^-$ with hidden charm but strangeness $S=-1$ were predicted, mostly made of  $\bar D^* \Xi_c$ or $\bar D^* \Xi^\prime_c$, decaying into $J/\psi \Lambda$.  In view of this, the decay of $\Xi^-_b$  into $J/\psi K^- \Lambda$ was suggested in \cite{cascadeb} as a suitable reaction to find a hidden charm strange state. Predictions for the $\bar K \Lambda$ and  $J/\psi \Lambda$ mass distributions were done, and, playing with uncertainties, it was shown that a clear peak in the $J/\psi \Lambda$ mass distribution should show up. This  reaction is presently being considered by the LHCb collaboration. However, since there is a much smaller statistics in the production of $\Xi^-_b$ than that of $\Lambda_b$ \cite{stone} , it is interesting to explore alternative reactions to observe this strangeness $S=-1$ hidden-charm pentaquark. In the present paper we suggest to employ the $ J/\psi \,\eta \Lambda$  decay mode of the $\Lambda_b$. Since the $\eta \Lambda$ pair is populated with a weight $\sqrt{2}/3$ relative to the $K^- p$ pair in the primary $\Lambda_b^0\to J/\psi M B$ reaction \cite{rocamai}, the $\Lambda_b \to J/\psi\, \eta \Lambda$  decay rate should be similar as that found for $J/\psi K^- p$ final states in the study of the non-strange pentaquark, and the new strange state should be looked for in the  $J/\psi \Lambda$ mass distribution instead of the $J/\psi p$ one.  We note that the possible existence of an strange $S=-1$ pentaquark partner was also studied in \cite{Lu:2016roh} from the non-strange decay mode $\Lambda_b \to J/\psi K^0 \Lambda$, which is one of the coupled channels of the decay $\Lambda_b \to J/\psi \pi^- p$ from which, even if it is more Cabbibo suppressed than the $\Lambda_b \to J/\psi K^- p$ process, a possible signal of the $P_c(4450)$ may also have been seen \cite{Aaij:2015fea,Wang:2015pcn}. The study of \cite{Lu:2016roh} explored the effect of different weak decay amplitudes to produce either a $J^P=1/2^-$ or a $J^P=3/2^-$ strange pentaquark. In this work, we will also take these possibilities into account. 
  
This paper is organized as follows. In Sect.~\ref{sec:formalism} we present our formalism for the $\Lambda_b \to J/\psi \,\eta \Lambda$ decay, describing the weak transition process and the implementation of final state interactions. Our results are shown in Sect.~\ref{sec:results}, where the spectra of both $\eta \Lambda$ and $J/\psi \Lambda$ states can be seen and a discussion on their dependence on various uncertain parameters of our model can be found.
We shall show that, even within uncertainties, the signal for a strangeness $S=-1$ hidden charm pentaquark remains as a clear peak in the $J/\psi \Lambda$ mass distribution. Our conclusions are summarized in Sect.~\ref{sec:conclusions}.

\section{Formalism}

\label{sec:formalism}

\subsection{The \mbox{\boldmath{$\Lambda_b \to J/\psi \, \eta \Lambda$}} decay process}
\label{subsec:form}
\begin{figure}[!htb]
 \includegraphics[width=\linewidth]{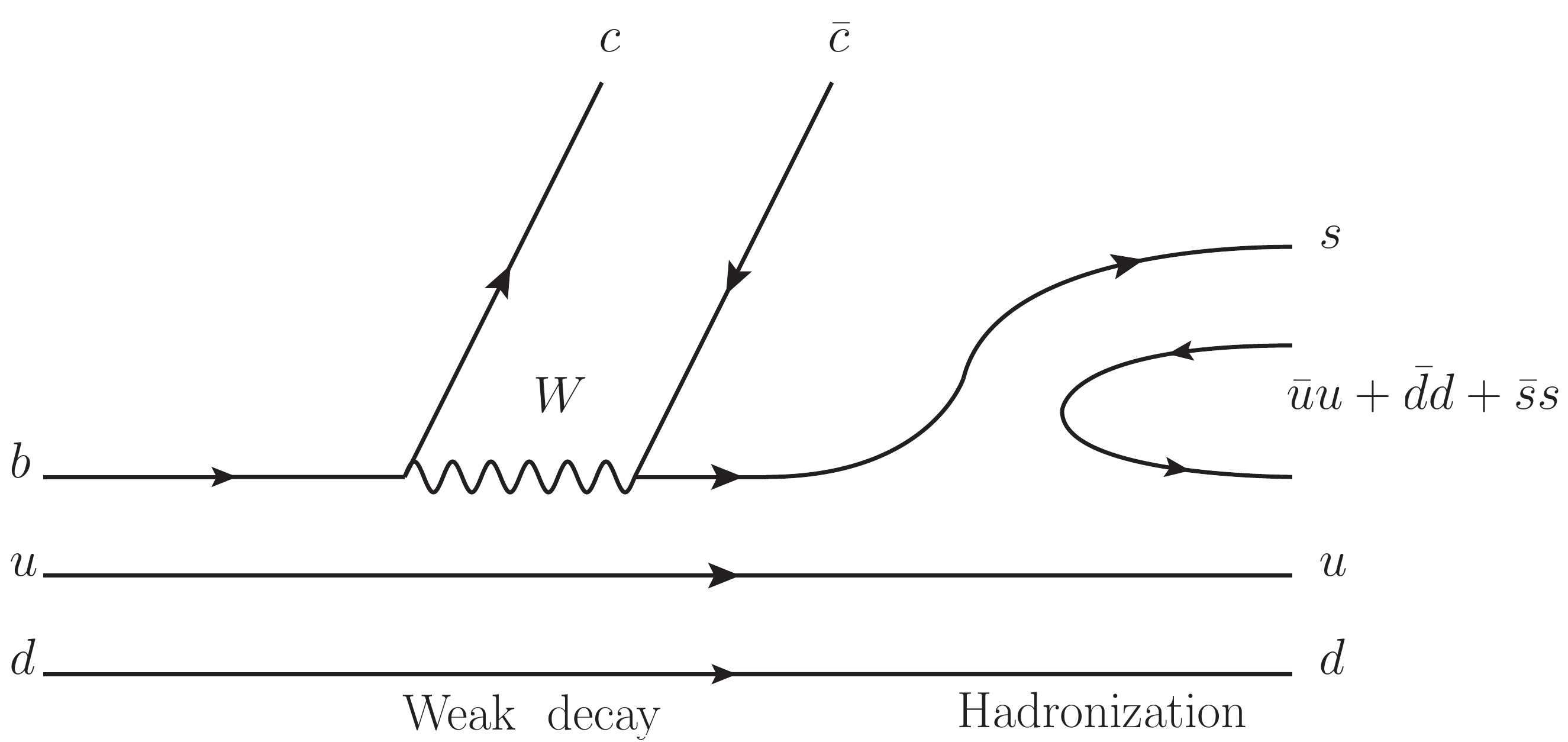}
\caption{Diagram describing the weak decay 
$\Lambda_b$ into the $J/\psi$ and a meson-baryon pair formed
through a hadronization mechanism.}\label{fig:decay}
\end{figure}

The study of the $\Lambda_b \to J/\psi \, \eta \Lambda$ decay follows the same approach as that presented in \cite{rocamai} for $\Lambda_b \to J/\psi ~K^-~ p$.  At quark level, both processes are identical and proceed through the transition diagram depicted in Fig.~\ref{fig:decay}, where we can see the $W$-exchange weak process transforming the $b$ quark into $c{\bar c} s$, followed by the hadronization of a pair of quarks which eventually produces a meson and a baryon, in addition to the $J/\psi$. The process depicted in Fig.~\ref{fig:decay} assumes that the elementary weak decay involves only the $b$ quark of the $\Lambda_b$, while the $u$ and $d$ quarks remain as spectators, the reason being that one expects one-body operators in a microscopical evaluation to have  larger strength than two- or multi-body operators. According to this assumption, since the $\Lambda_b$ has isospin $I=0$, so does the spectator $ud$ pair, which, combined with the $s$ quark after the weak decay, can only form $I=0$ $\Lambda$ states. The findings of the experimental analysis of Ref.~\cite{Aaij:2015tga} clearly support this hypothesis. 

For the hadronization process we introduce a $\bar{q}q$ pair between two quarks with the quantum numbers of the vacuum, $\bar{u}u + \bar{d}d + \bar{s}s$. The dominant contribution of the hadronization preserves the spectator role of the $ud$ pair, which ends up into the final baryon, and requires the involvement of the $s$ quark, which ends up into the final meson. Any other topology that would bring the $u$ or $d$ quark into the final meson requires a large momentum transfer that supresses the mechanism. If, in addition, we wish to have the meson-baryon pair in s-wave, it will have negative parity, forcing the $s$ quark prior to hadronization to have also this parity and thus be in an excited state. Since in the final $K^-$ or $\eta$ mesons the $s$ quark is in its ground state, this also implies that the $s$ quark produced immediately after the weak process must participate actively in the process of hadronization, which proceeds as shown in Fig.~\ref{fig:decay}. A further discussion on the reduced size of other alternative mechanisms can be found in \cite{miyahara}.

The technical way to implement the hadronization and produce meson-baryon pairs in the final state follows the same steps as in \cite{liang,dai,xie} for meson decays and in \cite{rocamai,feijoomagas} for the $\Lambda_b$ decay. The flavor decomposition of the $\Lambda_b$ state is:
\begin{equation}
 |\Lambda_b\rangle=\frac{1}{\sqrt{2}}|b(ud-du)\rangle \ ,
\end{equation}
which becomes, after the weak process
\begin{equation}
 |H\rangle=\frac{1}{\sqrt{2}}|s(ud-du)\rangle \ ,
\end{equation}
or,  upon hadronization,
\begin{equation}
 |H\rangle=\frac{1}{\sqrt{2}}|s(\bar{u} u+\bar{d}d+\bar{s}s)(ud-du)\rangle \ ,
\end{equation}
which can be written in terms of the $q\bar{q}$ matrix $P$, as
\begin{equation}
 |H\rangle=\frac{1}{\sqrt{2}}\sum_{i=1}^3|P_{3i}q_i(ud-du)\rangle \ ,
\end{equation}
where 
\begin{equation}
P=\left(\begin{array}{ccc}
            u\bar{u} & u\bar{d} & u\bar{s}\\
            d\bar{u} & d\bar{d} & d\bar{s}\\
            s\bar{u} & s\bar{d} & s\bar{s}
            \end{array}\right)  ~~\mbox{and}~~ q=\left(\begin{array}{c}
                                                                      u\\
                                                                      d\\
                                                                      s
                                                                      \end{array}\right)  \ .
\end{equation}
Writing the matrix $P$ in terms of the meson states, $P \to \phi$, where the $\eta, \eta'$ mixing \cite{bramon} has been assumed, 
\begin{equation}
\phi=\left(\begin{array}{ccc}\frac{\pi^0}{\sqrt{2}}+\frac{\eta}{\sqrt{3}}+\frac{\eta'}{\sqrt{6}} & \pi^+ & K^+\\
\pi^- & -\frac{\pi^0}{\sqrt{2}}+\frac{\eta}{\sqrt{3}}+\frac{\eta'}{\sqrt{6}} & K^0\\
K^- & \bar{K}^0 & -\frac{\eta}{\sqrt{3}}+\frac{2\eta'}{\sqrt{6}}\end{array}\right)\,,
\end{equation}      
the hadronized state becomes:
\begin{eqnarray}
|H\rangle=\frac{1}{\sqrt{2}}\Big( \Big. && K^-u(ud-du)+\bar{K}^0d(ud-du) \nonumber \\
&&\left.+\frac{1}{\sqrt{3}}\left(-\eta+\sqrt{2}\eta'\right)s(ud-du)\right)
\end{eqnarray}

By the former equation one obtains the mixed antisymmetric representation of the octet of baryons and taking the results of \cite{Close:1979bt} (see also \cite{rocamai}) one finds the final representation
\begin{equation}
|H\rangle=|K^-p\rangle+|\bar{K}^0n\rangle-\frac{\sqrt{2}}{3}|\eta\Lambda\rangle
\label{eq:primaryprod}
\end{equation}
where we have omitted the $|\eta'\Lambda\rangle$ contribution because of the large mass of the $\eta'$ meson \cite{rocamai}.

\begin{figure}[!htb]
 \includegraphics[width=\linewidth]{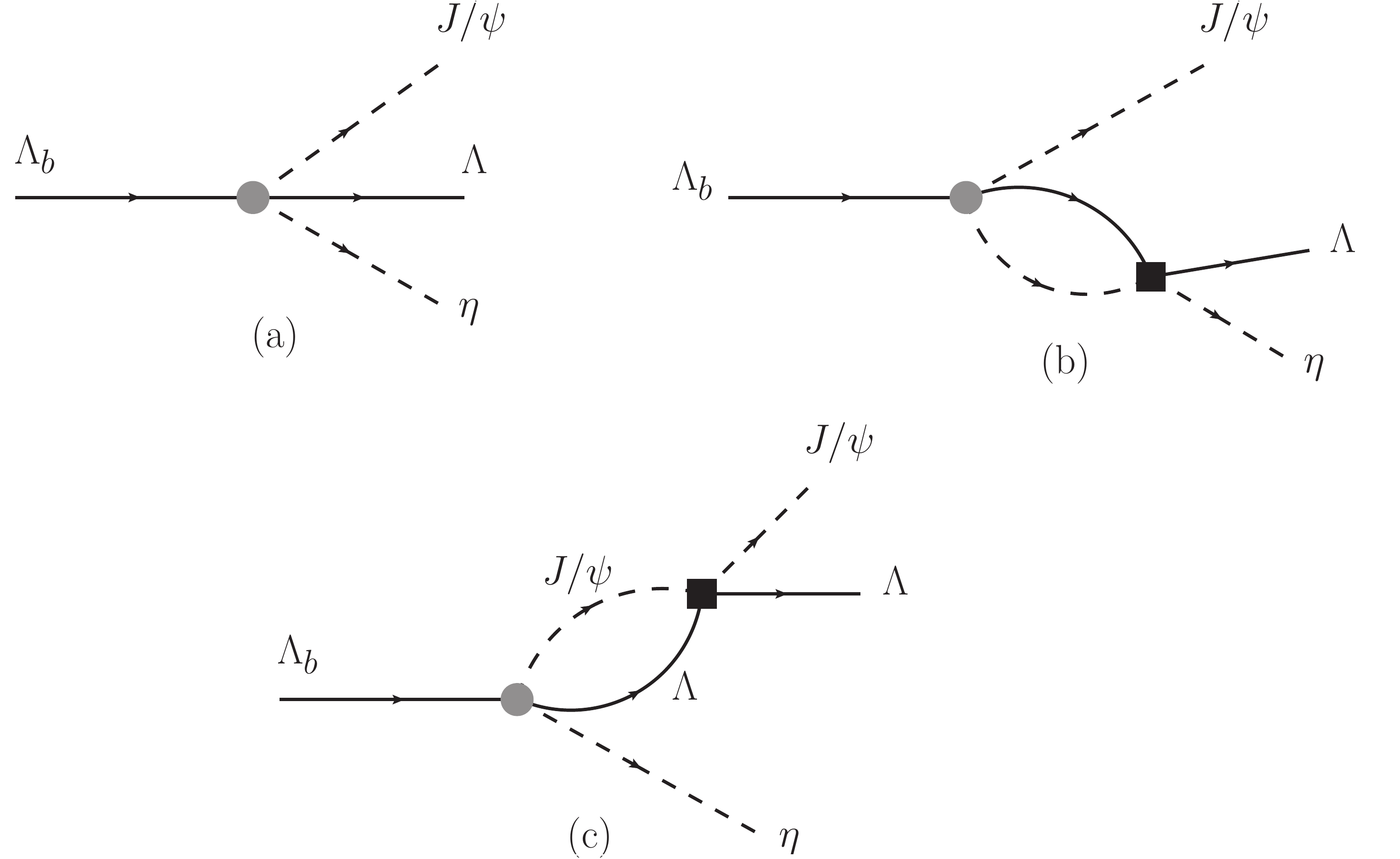}
\caption{Diagrammatic representation of the decay amplitude for $\Lambda_b \to J/\psi \,\eta\Lambda$: a) tree level, b) the $\eta\Lambda$ production through the coupled channel interaction of the initially produced $\eta\Lambda$ and $\bar{K}N$ meson-baryon pairs, c) $J/\psi\Lambda \to J/\psi\Lambda$ interaction.}\label{fig:diagrams}
\end{figure}

The final step consists in taking into account the final state interaction of the meson-baryon pairs. The amplitude for the $\Lambda_b \to J/\psi \, \eta \Lambda$ decay will then be built from the diagrams of Fig.~\ref{fig:diagrams}, where we can see the direct tree-level process, depicted by diagram (a),  the 
final-state interaction contribution of the meson-baryon pair into $\eta \Lambda$ production (b), and the final-state  $J/\psi\Lambda \to J/\psi\Lambda$ interaction (c). The corresponding amplitude can be written as:
\begin{eqnarray}
\mathcal{M}(M_{\eta \Lambda},M_{J/\psi \Lambda}) = V_{p} \Big[ h_{\eta\Lambda}  \Big.  \mkern-10mu &&
+ \mkern-5mu \sum_i h_i G_i( M_{\eta\Lambda}) t_{i,\eta\Lambda}( M_{\eta\Lambda}) \nonumber \\
&& \mkern-150mu 
\Big. + \, h_{\eta\Lambda} G_{J/\psi \Lambda}(M_{J/\psi \Lambda}) \, t_{J/\psi \Lambda,J/\psi \Lambda}(M_{J/\psi \Lambda} )\Big] ,
\label{eq:amplitude}
\end{eqnarray}
where the weights $h_i$, obtained from Eq.~(\ref{eq:primaryprod}), are:
\begin{eqnarray}
&h_{\pi^0\Sigma^0}=h_{\pi^+\Sigma^-}=h_{\pi^-\Sigma^+}=0\,,~h_{\eta\Lambda}=
-\frac{\sqrt{2}}{3}\,,\\
&h_{K^-p}=h_{\bar K^0n}=1\,,~h_{K^+\Xi^-}=h_{K^0\Xi^0}=0\ ,
\end{eqnarray}
and where $G_i$, with $i={K^-p, \bar{K}^0 n, \eta\Lambda}$, denotes the meson-baryon loop function, chosen in 
accordance with the model for the scattering matrix $t_{i,\eta\Lambda}$ \cite{Feijoo:2015yja}. Similarly, we take the loop function $G_{J/\psi \Lambda}$ employed in the model of \cite{Wu:2010jy,Wu:2010vk} on which, as we will show below, we base our prescription for $t_{J/\psi \Lambda,J/\psi \Lambda}$.  The factor $V_p$, which includes the common dynamics of the production of the different pairs, is unknown and we take it as constant, see  Ref.~\cite{feijoomagas} for a more detailed argumentation. 

\begin{figure}[!htb]
\center
 \includegraphics[width=0.6\linewidth]{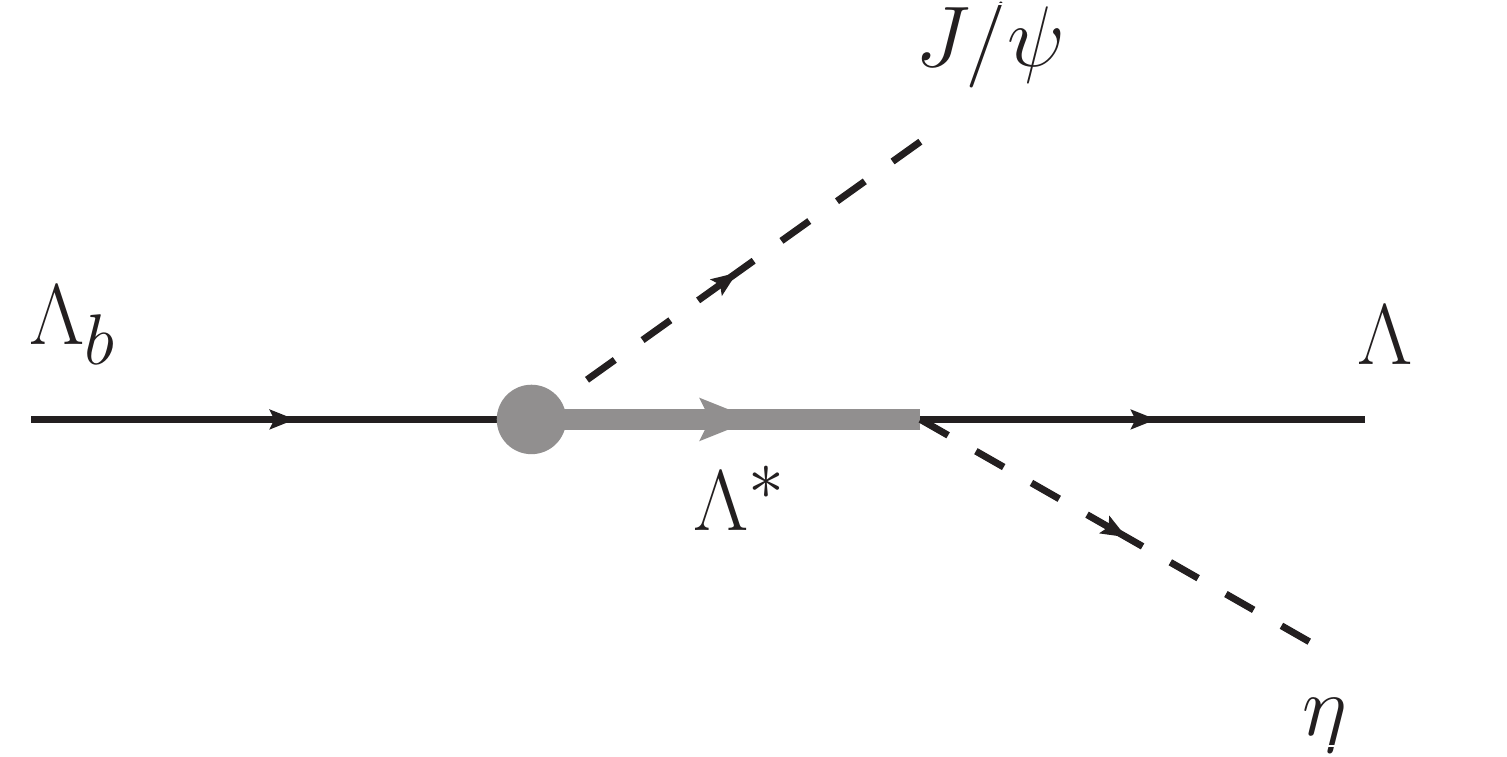}
\caption{Diagrammatic representation of an s-wave resonance contribution to the $\Lambda_b \to J/\psi \,\eta\Lambda$ decay amplitude}\label{fig:reson}
\end{figure}

At this point it is worth mentioning that the model for the final state interaction in the $\eta \Lambda$ channel, which is briefly described in the next section, generates some resonances dynamically, like the $\Lambda(1405)$ or the $\Lambda(1670)$, that are either below or at the edge of the threshold of $\eta\Lambda$ invariant masses $M_{\eta\Lambda}$ accessible from the decay of the  $\Lambda_b$. We therefore would like to explore the possibility of adding to the amplitude the explicit contribution of  some $\Lambda^*$ which lies in the relevant $M_{\eta\Lambda}$ region, essentially between 1700 MeV and 2500 MeV, and might couple sensibly to $\eta\Lambda$ states, as represented diagrammatically in Fig.~\ref{fig:reson}.  One state with these characteristics is listed in the PDG compilation \cite{PDG}, the one star  $\Lambda(2000)$, having a width $\Gamma \sim 100-300$ MeV and a branching ratio to the decay into $\eta\Lambda$ of $(16\pm7)\%$. The recent unitary multichannel model for $\bar{K} N$ scattering, with parameters fitted to partial waves up to $J=7/2$ and up to 2.15 GeV of energy, also gives an s-wave $J^P=1/2^-$  $\Lambda$ state with similar mass and width properties \cite{Fernandez-Ramirez:2015tfa}.

\begin{figure}[!htb]
\center
 \includegraphics[width=0.8\linewidth]{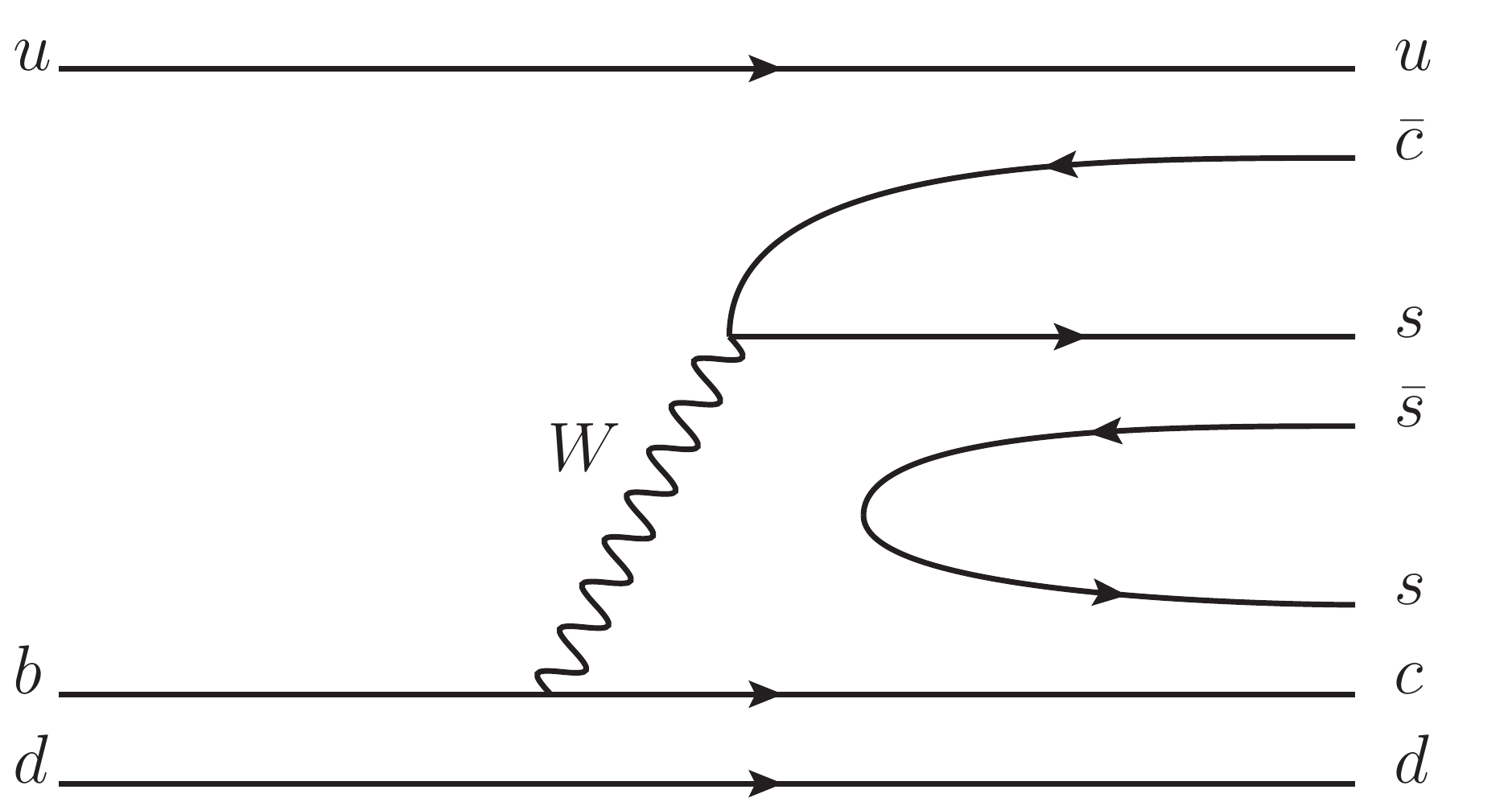}
\caption{Diagram describing the weak decay 
$\Lambda_b$ into the $\bar D^*$ and a $\eta\Xi^\prime_c$ pair.}
\label{fig:decay2}
\end{figure}

\begin{figure}[!htb]
 \includegraphics[width=\linewidth]{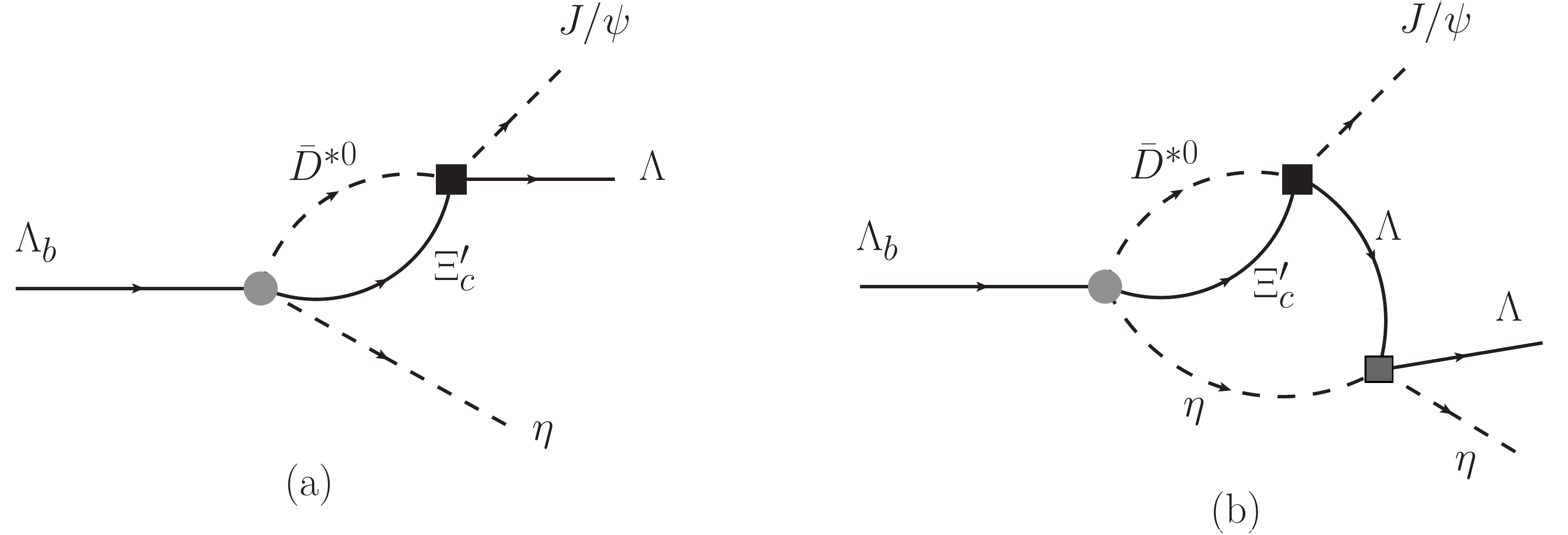}
\caption{Diagrammatic representation of the decay of the  $\Lambda_b$ into a virtual $\eta \bar{D}^{*0}\Xi_c^\prime $ intermediate state, followed by the  $\bar{D}^{*0}\Xi_c^\prime  \to J/\psi \Lambda$ conversion process, (a), and implementing also the final state interaction of the final $\eta\Lambda$ pair, (b).}
\label{fig:diagram5}
\end{figure}

We note that, since our model will rely on the strange pentaquark predicted in Refs.~\cite{Wu:2010jy,Wu:2010vk} at an energy around 4450 MeV, which couples strongly to $\bar{D}^{*0}\Xi^\prime_c$ states, one should consider the possibility that 
the influence of this resonance in the final $J/\psi \Lambda$ mass distribution could also be due to the creation of  a virtual $\bar{D}^*\eta \Xi^\prime_c $  state in a first step of the $\Lambda_b$ decay, through the mechanism of Fig.~\ref{fig:decay2}, followed by multiple interactions to generate the resonance, which would eventually decay into a  $J/\psi \Lambda$ pair in the final state,  represented by the diagrams of Fig.~\ref{fig:diagram5}. However, this configuration requires a different topology, as seen in Fig.~\ref{fig:decay2}, in which the $ud$ quarks of the $\Lambda_b$ do not act as a coupled spectator pair. Although it is hard to quantify the size of the amplitude of Fig.~\ref{fig:decay} with respect to that of  Fig.~\ref{fig:decay2}, the fact that in this later case one of the spectator quarks ends up in the charmed meson and the other one goes to the baryon makes us believe that the corresponding amplitude will be reduced. We will therefore assume the dominance of the mechanism of Fig.~\ref{fig:decay} over that of Fig.~\ref{fig:decay2} by a factor of two or more and will give predictions for different relative signs of the two processes. We note that the lowest order contribution to the $\Lambda_b \to J/\psi\,\eta \Lambda$ decay induced by virtual $\bar{D}^*\eta\Xi^\prime_c$ states is the amplitude of Fig.~\ref{fig:diagram5} (a). We have checked, by explicit numerical evaluation, that the next-order contribution of Fig.~\ref{fig:diagram5} (b), involving the additional final state interaction of the $\eta\Lambda$ pair, gives a negligible correction, hence it will be ignored in the results presented in Sect.~\ref{sec:results}.

Adding the s-wave resonant contribution of Fig.~Ê\ref{fig:reson} and the process initiated by an intermediate $\bar{D}^*\eta\Xi^\prime_c $  state followed by final state interactions leading to a $J/\psi \Lambda$ pair and an $\eta$ meson represented by the diagram of Fig.~\ref{fig:diagram5} (a), the final amplitude for $\Lambda_b \to J/\psi\,\eta\Lambda$ decay, producing a strange pentaquark with $J^P=1/2^-$ becomes:
\begin{eqnarray}
\mathcal{M}(M_{\eta \Lambda},M_{J/\psi \Lambda}) = V_{p} \Big[ h_{\eta\Lambda}  \Big.  \mkern-10mu &&
+ \mkern-5mu \sum_i h_i G_i( M_{\eta\Lambda}) t_{i,\eta\Lambda}( M_{\eta\Lambda}) \nonumber \\
&& \mkern-105mu 
+ \, h_{\eta\Lambda} G_{J/\psi \Lambda}(M_{J/\psi \Lambda}) \, t_{J/\psi \Lambda,J/\psi \Lambda}(M_{J/\psi \Lambda} )\nonumber \\
&& \mkern-105mu  
 +  \, \, \beta \, \, G_{\bar{D}^*\Xi^\prime_c }(M_{\bar{D}^*\Xi^\prime_c}) \,  t_{\bar{D}^*\Xi^\prime_c,J/\psi \Lambda}(M_{J/\psi \Lambda}) \nonumber \\
&& \mkern-105mu \left. + \, \alpha \, \frac{M_{\Lambda^*(2000)}}{M_{\eta\Lambda}-M_{\Lambda^*(2000)}+ {\rm i}\, \frac{\Gamma_{\Lambda^*(2000)}}{2}}
\right]\,,
\label{eq:amplitude2}
\end{eqnarray}
where $\alpha$ is a dimensionless parameter that determines the strength of the s-wave resonant mechanism, while $\beta$ controls the strength of $\Lambda_b$ decaying virtually into $\bar{D}^*\eta\Xi^\prime_c $, relative to its decay into $J/\psi\, \eta  \Lambda$.  

The amplitudes of Eqs.~(\ref{eq:amplitude}) and  (\ref{eq:amplitude2}) come with the matrix element $\langle m_\Lambda \mid \vec{\sigma}\vec{\epsilon} \mid m_{\Lambda_b} \rangle $, tied to the s-wave character assumed for the weak decay vertex and accounting for the spin $1/2$ of the decaying $\Lambda_b$, and the spins 1/2 and 1 of the emitted $\Lambda$ and $J/\psi$ meson, respectively. Moreover,
as will be recalled in the next subsections, the $J/\psi\Lambda$ (and $\eta\Lambda$) interaction models are also taken in s-wave, hence the spin values of the $J/\psi\Lambda$ pair could  in principle be $J=1/2$ or $J=3/2$. However, only the $J=1/2$ case is allowed to match the spin 1/2 of the decaying $\Lambda_b$. In fact the $\vec{\sigma}\vec{\epsilon}$
operator projects the $J/\psi\Lambda$ system into $J=1/2$ as shown in \cite{Garzon:2012np}. 

Our general strategy is to assume that the decay process proceeds involving the smallest possible angular momentum at the vertices. Therefore, in order to produce the strange pentaquark with $J=3/2$ it is necessary to implement at least a p-wave contribution in the weak decay mechanism. A p-wave operator of the form
\begin{equation}
T^{\rm p-wave}_{\rm tree} = {\rm i} B \varepsilon_{ijk} \sigma_k q_i \epsilon_j 
\label{eq:pwave}
\end{equation}
was considered in \cite{Lu:2016roh}, where it was also shown that, if the peak seen in \cite{Aaij:2015tga} corresponds to the molecular states with $J^P=3/2^-$ generated in \cite{Wu:2010jy,Wu:2010vk} from the scalar-vector meson interaction in s-wave, the momentum $\vec{q}$ in the former equation must
be that of the  $\eta$-meson, which is taken in the rest frame of the $\eta\Lambda$ system. A decomposition of the p-wave vertex in terms of two operators, 
\begin{eqnarray}
\label{eq:projectors}
\hat{S}_{3/2}&=&\langle m_\Lambda \mid q_j \epsilon_j + \frac{\rm i}{2} \varepsilon_{ijk} \sigma_k q_i \epsilon_j \mid m_{\Lambda_b} \rangle \nonumber \\
\hat{S}_{1/2}&=&\langle m_\Lambda \mid q_j \epsilon_j - {\rm i} \varepsilon_{ijk} \sigma_k q_i \epsilon_j \mid m_{\Lambda_b} \rangle \ ,
\end{eqnarray}
that project, respectively, over the spin $J=3/2$ and $J=1/2$ of the two-body $J/\psi\Lambda$ system, was also given in \cite{Lu:2016roh}:
\begin{equation}
T^{\rm p-wave}_{\rm tree} = {\rm i} B \varepsilon_{ijk} \sigma_k q_i \epsilon_j  = \frac{2}{3} B \, \hat{S}_{3/2} - 
 \frac{2}{3} B \, \hat{S}_{1/2} \  .
\label{eq:descomposition}
\end{equation}
The $J=3/2$ pentaquark will then be generated by the final state interaction of the $J/\psi\Lambda$ pair initiated by the p-wave decay vertex of Eq.~(\ref{eq:pwave}), in a  process of the type represented by the diagram of Fig.~\ref{fig:diagrams}(c), or also from the virtual excitation of intermediate $\bar{D}^*\eta \Xi^\prime_c $ states followed by multiple interactions leading to $J/\psi\Lambda$ pairs in the final state, as seen in Fig.~\ref{fig:diagram5}(a). Note that in this later case the p-wave decay vertex will have the same structure as that of Eq.~(\ref{eq:pwave}), but with a strength constant $B^\prime$, and will not act at tree-level.

Considering the s-wave and p-wave contributions, the amplitude that allows for the appearance of a pentaquark with $J=3/2$ is then given by:
\begin{eqnarray}
&&\mathcal{M}(M_{\eta \Lambda},M_{J/\psi \Lambda}) = \nonumber \\
&& \mkern-5mu   V_{p} \left[ h_{\eta\Lambda} + \sum_i h_i G_i( M_{\eta\Lambda}) t_{i,\eta\Lambda}( M_{\eta\Lambda}) \right] 
\langle m_\Lambda \mid \vec{\sigma}\vec{\epsilon} \mid m_{\Lambda_b} \rangle   \nonumber \\
&& \mkern-5mu 
+\frac{2}{3} B \, \hat{S}_{3/2} - 
 \frac{2}{3} B \, \hat{S}_{1/2} \nonumber \\
&& \mkern-5mu+\frac{2}{3} B\, G_{J/\psi \Lambda}(M_{J/\psi \Lambda}) \, t_{J/\psi \Lambda,J/\psi \Lambda}(M_{J/\psi \Lambda} )\, \hat{S}_{3/2} \ \nonumber \\
&& \mkern-5mu+\frac{2}{3} B^\prime\, G_{\bar{D}^*\Xi^\prime_c }(M_{\bar{D}^*\Xi^\prime_c}) \,  t_{\bar{D}^*\Xi^\prime_c,J/\psi \Lambda}(M_{J/\psi \Lambda}) \, \hat{S}_{3/2}  \ ,
\label{eq:amp_pwave}
\end{eqnarray}
where the term proportional to $V_p$ stands for the contribution of the s-wave weak decay amplitude\footnote{Note that we have omitted here the contribution of an explicit $\Lambda(2000)$ resonance, since its effect does not bring any qualitative changes in the  $J/\psi\Lambda$ pair spectrum, as will be shown in the Results section.},  the next row contains the contribution of the p-wave tree level term and the last two rows correspond to the final state interaction contributions that generate the $J=3/2$ pentaquark initiated by $J/\psi\Lambda$ states (proportional to $B$) or by $\bar{D}^*\Xi^\prime_c$ states (proportional to $B^\prime$). The former equation can be cast schematically as:
\begin{equation}
\mathcal{M} = C_1 \, \langle m_\Lambda \mid \vec{\sigma}\vec{\epsilon} \mid m_{\Lambda_b} \rangle   + C_2 \, \hat{S}_{3/2} + C_3 \hat{S}_{1/2}  \ .
\end{equation}

Finally, the double differential cross-section for the $\Lambda_b \to J/\psi \, \eta \Lambda$ decay process reads \cite{rocamai}:
\begin{eqnarray}
&&\frac{d^2\Gamma}{dM_{\eta \Lambda}dM_{J/\psi \Lambda}}  = \nonumber \\
&&
\frac{1}{{(2\pi)}^3}\frac{4M_{\Lambda_b}M_{\Lambda}}{32M_{\Lambda_b}^3}\overline{\sum}| \mathcal{M}(M_{\eta \Lambda},M_{J/\psi \Lambda})|^2 2 M_{\eta \Lambda} 2 M_{J/\psi \Lambda}  \  , \nonumber \\
\label{eq:double_diff_cross}
\end{eqnarray}
where, after performing the sum over final spins and polarizations and the average over initial spins (see appendix in \cite{Lu:2016roh}), one has:
\begin{equation}
\overline{\sum}| \mathcal{M}(M_{\eta \Lambda},M_{J/\psi \Lambda})|^2 =
3 | \mathcal{M}(M_{\eta \Lambda},M_{J/\psi \Lambda})|^2 \ ,
\end{equation}
with $\mathcal{M}$ being that of Eqs.~(\ref{eq:amplitude}) or (\ref{eq:amplitude2}),
corresponding to an s-wave weak vertex and, hence, producing a pentaquark with $J=1/2$, or
\begin{equation}
\overline{\sum}| \mathcal{M}(M_{\eta \Lambda},M_{J/\psi \Lambda})|^2 =
3 | C_1 |^2 + \frac{3}{2} \vec{q}^2 |C_2|^2 + 3 \vec{q}^2 |C_3|^2 \ ,
\end{equation}
with $\mathcal{M}$ being that of Eq.~(\ref{eq:amp_pwave}),
corresponding to a weak vertex that also has a p-wave term and, hence, making the production of a pentaquark with $J=3/2$ possible.

Fixing the invariant mass $M_{J/\psi \Lambda}$, one can integrate over $M_{\eta \Lambda}$ in order to obtain $d\Gamma/dM_{J/\psi \Lambda}$. In this case, the limits are given by:
\begin{eqnarray}
\left( M_{\eta \Lambda}^2 \right)_{\rm max} &=&{\left(E^*_{\Lambda}+E^*_{\eta}\right)}^2 \nonumber \\
&&-{\left(\sqrt{{E^*_{\Lambda}}^2-M^2_{\Lambda}}-\sqrt{{E^*_{\eta}}^2-m^2_{\eta}}\right)}^2
\end{eqnarray}
and
\begin{eqnarray}
\left( M_{\eta \Lambda}^2 \right)_{\rm min}&=&{\left(E^*_{\Lambda}+E^*_{\eta}\right)}^2 \nonumber \\
&&-{\left(\sqrt{{E^*_{\Lambda}}^2-M^2_{\Lambda}}+\sqrt{{E^*_{\eta}}^2-m^2_{\eta}}\right)}^2 \ ,
\end{eqnarray}
where
\begin{equation}
E^*_{\Lambda}=\frac{M_{J/\psi \Lambda}^2-m_{J/\psi}^2+M^2_{\Lambda}}{2M_{J/\psi \Lambda}} \ ,
\end{equation}
\begin{equation}
E^*_{\eta}=\frac{M_{\Lambda_b}^2-M_{J/\psi \Lambda}^2-m^2_{\eta}}{2M_{J/\psi \Lambda}} \ .
\end{equation}
Similar formulas are obtained when fixing the invariant mass $M_{\eta \Lambda}$ and integrating over $M_{J/\psi \Lambda}$ to obtain $d\Gamma/dM_{\eta \Lambda}$:
\begin{eqnarray}
\left( M_{J/\psi \Lambda}^2 \right)&&_{\rm max}={\left(E^*_{\Lambda}+E^*_{J/\psi}\right)}^2 \nonumber \\
&&-{\left(\sqrt{{E^*_{\Lambda}}^2-M^2_{\Lambda}}-\sqrt{{E^*_{J/\psi}}^2-m^2_{J/\psi}}\right)}^2 ,
\end{eqnarray}
\begin{eqnarray}
\left( M_{J/\psi \Lambda}^2 \right)&&_{\rm min} ={\left(E^*_{\Lambda}+E^*_{J/\psi}\right)}^2 \nonumber \\
&&-{\left(\sqrt{{E^*_{\Lambda}}^2-M^2_{\Lambda}}+\sqrt{{E^*_{J/\psi}}^2-m^2_{J/\psi}}\right)}^2 ,
\end{eqnarray}
where
\begin{equation}
E^*_{\Lambda}=\frac{M_{\eta \Lambda}^2-m_{\eta}^2+M^2_{\Lambda}}{2M_{\eta \Lambda}} \ ,
\end{equation}
\begin{equation}
E^*_{J/\psi}=\frac{M_{\Lambda_b}^2-M_{\eta \Lambda}^2-m^2_{J/\psi}}{2M_{\eta \Lambda}} \ .
\end{equation}

\subsection{Final interaction models}
\label{subsec:models}
In this section we briefly describe the theoretical models employed to obtain the amplitudes $t_{i,\eta\Lambda}$,  $t_{J/\psi \Lambda, J/\psi \Lambda}$ and $t_{\bar{D}^*\Xi^\prime_c, J/\psi \Lambda}$, which account for the final state interaction effects.

The $S=-1$ meson-baryon amplitude with $\eta\Lambda$ in the final state appearing in diagram (b) of Fig.~ \ref{fig:diagrams} is determined from the coupled-channel unitary model of Ref.~\cite{Feijoo:2015yja}, developed with the aim of improving upon the knowledge of the chiral interation at next-to-leading order (NLO). The parameters of the model were fitted to a large set of experimental scattering data \cite{exp_data_scattering}, as well as to branching ratios at threshold \cite{exp_data_ratios}, and to the precise SIDDHARTA value of the energy shift and width of kaonic hidrogen \cite{Bazzi:2011zj}. Differently to other works, as e.g. \cite{ollerulf,Borasoy:2006sr,hyodonew}, the model was also constrained to reproduce the $K^- p\to K^+\Xi^-, K^0\Xi^0$ reactions, since they  are especially sensitive to the NLO terms. The work of Ref.~\cite{Feijoo:2015yja} also 
investigated the influence of high spin hyperon resonances on the $K^- p \to K\Xi$ amplitudes, finding that
the resonant terms helped in improving the description of the scattering data and produced more precise values of the low energy constants of the chiral unitary model.

More especifically, the meson-baryon amplitudes of Ref.~\cite{Feijoo:2015yja} are built from a kernel obtained from the SU(3) chiral Lagrangian up to NLO:
\begin{equation}
v_{ij}=v^{\scriptscriptstyle WT}_{ij}+v^{\scriptscriptstyle NLO}_{ij} 
\label{eq:kernel}
\end{equation}
where
\begin{equation}
v^{\scriptscriptstyle WT}_{ij}=
 - \frac{C_{i j}(2\sqrt{s} - M_{i}-M_{j})}{4 f^2}\! N_{i} N_{j}
\end{equation}
and
\begin{equation}
v^{\scriptscriptstyle NLO}_{ij} =\frac{D_{ij}-2(k_{i,\mu} k_j^{\mu})L_{ij}}{f^2}\! N_{i} N_{j}\ ,
\end{equation}
with
\begin{equation}
N_{i}=\sqrt{\frac{M_i+E_i}{2M_i}},\,\, N_{j}=\sqrt{\frac{M_j+E_j}{2M_j}} \nonumber \ .
\end{equation}\
The indices $i,j$ stand for any of the ten meson-baryon channels in the neutral $S=-1$ sector: $K^-p$, $\bar{K}^0 n$, $\pi^0\Lambda$, $\pi^0\Sigma^0$, $\pi^-\Sigma^+$, $\pi^+\Sigma^-$, $\eta\Lambda$, $\eta\Sigma^0$, $K^+\Xi^-$ and $K^0\Xi^0$, while $M_i,M_j$ and $E_i,E_j$ are the masses and energies, respectively, of the baryons involved in the transition,
and $k_{i,\mu}, k_j^{\mu}$ are the corresponding meson four-momenta. The lagrangian is written in terms of SU(3) coefficients $C_{ij}$, the pion decay constant $f$ and other low energy constants embedded in the matrices  
$D_{ij}$ and $L_{ij}$ of the NLO term, which can be found, for example, in the appendices of Ref.~\cite{Feijoo:2015yja}.  

Chiral unitary amplitudes are obtained by solving the Bethe-Salpeter equation in its on-shell factorized form:
\begin{equation}
t_{ij} =v_{ij}+v_{il} G_l t_{lj}  \ ,
\label{LS}
\end{equation} 
where the meson-baryon loop function $G_l$ is obtained employing dimensional regularization 
\begin{eqnarray} \label{Loop_integral}
G_l &=&{\rm i}\int \frac{d^4q_l}{{(2\pi)}^4}\frac{2M_l}{{(P-q_l)}^2-M_l^2+{\rm i}\epsilon}\frac{1}{q_l^2-m_l^2+{\rm i}\epsilon} \nonumber \\
& = &\frac{2M_l}{(4\pi)^2} \Bigg \lbrace a_l+\ln\frac{M_l^2}{\mu^2}+\frac{m_l^2-M_l^2+s}{2s}\ln\frac{m_l^2}{M_l^2} + \nonumber \\ 
 &     &\frac{q_{\rm cm}}{\sqrt{s}}\ln\left[\frac{(s+2\sqrt{s}q_{\rm cm})^2-(M_l^2-m_l^2)^2}{(s-2\sqrt{s}q_{\rm cm})^2-(M_l^2-m_l^2)^2}\right]\Bigg \rbrace ,  
\end{eqnarray}
where $M_l$ and $m_l$ are the baryon and meson masses of the $``l"$ channel, the regularization scale $\mu$ is taken to be 1~GeV , and $a_l$ are subtraction constants, which, together with the low energy parameters of the lagrangian, were determined from  fits to data performed in Ref.~\cite{Feijoo:2015yja}. We will employ the set of parameters corresponding to the model named ``NLO*" there, which will be referred to as Model 1 here.  These results will be compared to those obtained with another set of parameters --named  ``WT (no $K \Xi$)" in Ref.~\cite{Feijoo:2015yja} and referred to as WT in the present paper-- obtained from a fit that employs the lowest order Weinberg-Tomozawa
(WT) term without taking into account the
experimental data corresponding to the $K \Xi$ channels, as most of the works in this field. 

The chiral approach was complemented with the explicit inclusion, in the $K^-p \to K^0\Xi^0,K^+ \Xi^-$ amplitudes, of two high spin resonances, $\Sigma(2030)$ and  $\Sigma(2250)$, selected from the possible candidates listed in the PDG \cite{PDG} and in accordance to other resonance-based models \cite{Sharov:2011xq,jackson}. The spin and parity $J^\pi =7/2^+$ of the $\Sigma(2030)$ are well established. Those of the $\Sigma(2250)$ are not known,  but the choice
$J^\pi =5/2^-$ was adopted out of the two most probable assignments, $5/2^-$ or $9/2^-$. The fit of the model that includes the resonances, named ``NLO+RES" in  \cite{Feijoo:2015yja} and Model 2 here, determines not only the low energy parameters and subtraction constants but also the couplings, masses, widths and form-factor cut-offs of the resonances. 
More details on the implementation of the resonant terms can be found in Ref.~\cite{Feijoo:2015yja}.

With respect to the final state interaction in the $J/\psi \Lambda$ sector, represented by the diagrams of Fig.~\ref{fig:diagrams}(c)  and Fig.~\ref{fig:diagram5}(a), we recall that two states with strangeness and hidden charm with $J^P={3/2}^-$ and $I=0$ were found in Refs.~\cite{Wu:2010jy,Wu:2010vk} as meson-baryon molecules, having pole positions $\sqrt{s}=4368-2.8{\rm i}$ and $\sqrt{s}=4547-6.4{\rm i}$ and coupling to $J/\psi \Lambda$ states with strength $\mid g_{J/\psi \Lambda}\mid =0.47$ and $0.61$, respectively. The magnitude of each of these cou\-plings is relatively small compared to the coupling of the pole to the main meson-baryon component, which for the lower energy pole is $\bar{D}^*\Xi_c$, with $\mid g_{\bar{D}^*\Xi_c}\mid=3.6$, while for the higher energy one is $\bar{D}^*\Xi_c^\prime$, with $\mid g_{\bar{D}^*\Xi_c^\prime}\mid=2.6$. In any case, $\mid g_{J/\psi \Lambda}\mid$ is large enough to create a peak in the mass distribution, as we shall see. As candidate for the strangeness $-1$ pentaquark, we will consider the state at higher energy since its mass is close to the non-strange pentaquark found in \cite{Aaij:2015tga}.
One must however accept that the mass obtained for these states has uncertainties since, unlike in other sectors, one does not have any experimental data to constrain the parameters of the theory. We therefore take the nominal value of about $M_R=4550$ MeV for the mass of the strange pentaquark and will explore the stability of our results to variations of this mass. We shall take $\Gamma_R=10$ MeV in agreement with the findings of \cite{Wu:2010jy,Wu:2010vk}. 
Our explorations are implemented employing the following Breit-Wigner representation for the $t_{J/\psi \Lambda, J/\psi \Lambda}$ and $t_{\bar{D}^*\Xi^\prime_c,J/\psi \Lambda}$ amplitudes
\begin{equation}
\label{eq:tjpsi}
t_{J/\psi \Lambda,J/\psi \Lambda}=\frac{g^2_{J/\psi \Lambda}}{M_{J/\psi \Lambda}-M_R+{\rm i}\, \frac{\Gamma_R}{2}} \ ,
\end{equation}
\begin{equation}
\label{eq:tdstxi}
t_{\bar{D}^*\Xi^\prime_c,J/\psi \Lambda}=\frac{g_{\bar{D}^*\Xi^\prime_c}\,g_{J/\psi \Lambda}}{M_{J/\psi \Lambda}-M_R+{\rm i}\, \frac{\Gamma_R}{2}} \  .
\end{equation}

Then the production of the resonance is done through the $J/\psi \Lambda \to J/\psi \Lambda$ and ${\bar{D}^*\Xi^\prime_c \to J/\psi \Lambda}$ amplitudes, pa\-ra\-me\-tri\-zed through the expressions given in Eqs.~(\ref{eq:tjpsi}) and ~(\ref{eq:tdstxi}), as seen in diagrams of Fig.~\ref{fig:diagrams}(c) and Fig.~\ref{fig:diagram5}(a), respectively, as well as in Eqs.~(\ref{eq:amplitude}), (\ref{eq:amplitude2}) or (\ref{eq:amp_pwave}). The values of the couplings are  $g_{J/\psi \Lambda}=-0.61-0.06{\rm i}$ and  $ g_{\bar{D}^*\Xi_c^\prime} = 2.61 -0.13{\rm i}$.
The loop functions $G_{J/\psi \Lambda}$ and $G_{\bar{D}^*\Xi^\prime_c}$  appearing in these equations are taken from \cite{Wu:2010jy,Wu:2010vk}, where a dimensional regularization method with a scale $\mu = 1000$ MeV was employed, using subtraction constants $a_{J/\psi \Lambda}=a_{\bar{D}^*\Xi^\prime_c}=-2.3$.

\section{Results}
\label{sec:results}

\begin{figure}[!htb]
\centering
  \includegraphics[width=\linewidth]{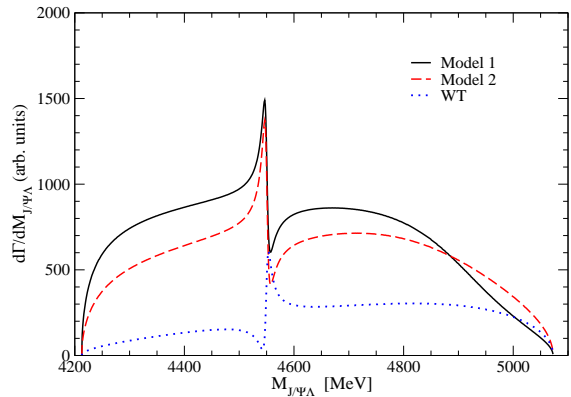}
\caption{(Color online) Invariant mass distributions of $J/\psi \Lambda$ states produced in the decay $\Lambda_b \to J/\psi ~ \eta \Lambda $, obtained for three models discussed in the text: one that considers only the WT term of the Lagrangian (dotted line) and two models, Model 1 (dashed line) and Model 2 (solid line) that incorporate also the next-to-leading order terms.}
  \label{fig:models_JPsi}
\end{figure}

We start this section by presenting, in Fig.~\ref{fig:models_JPsi}, the invariant mass distributions of $J/\psi \Lambda$ states produced in the decay $\Lambda_b \to J/\psi ~ \eta \Lambda $, obtained from the simpler s-wave weak decay approach of Eq.~(\ref{eq:amplitude}) and for three different models of the $S=-1$ $\eta\Lambda$ interaction \cite{Feijoo:2015yja}: one that only  considers the lowest-order WT term of the Lagrangian (dotted line) and two other models, Model 1 (dashed line) and Model 2 (solid line), that incorporate the next-to-leading order terms and, in the case of the later one, the effect of higher spin resonances.  Please note that although the Model 2 includes the additional contribution of the two high-spin resonances,  they do not contribute directly to the studied $I=0$ decay due to their $I=1$ nature, but their inclusion does modify the parameters of Model 2 with respect to those of Model 1.

 For the three models of the $\eta\Lambda$ interaction,
the peak of the pentaquark is clearly observed at 4550 MeV, the value of the mass $M_R$ employed in the parametrization of Eqs.~(\ref{eq:tjpsi}) and (\ref{eq:tdstxi}). However the overall strength is enhanced for the NLO models, which also show a different interference pattern with the non-resonant background to that of the lowest-order WT model. 

\begin{figure}[!htb]
\centering
  \includegraphics[width=\linewidth]{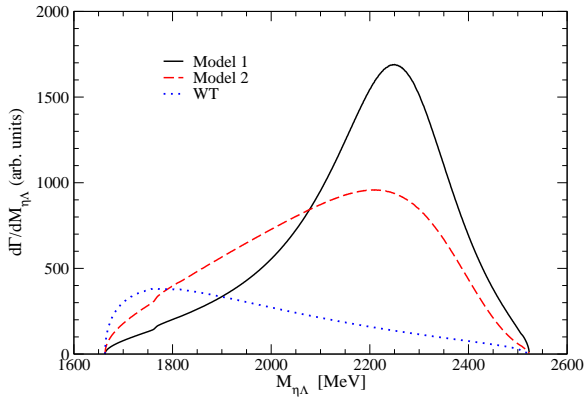}
\caption{(Color online) The same as Fig.~\ref{fig:models_JPsi} but for the invariant mass distributions of $\eta \Lambda$ states produced in the decay $\Lambda_b \to J/\psi ~ \eta \Lambda $.}
  \label{fig:models_eta}
\end{figure}

The invariant mass distribution of $\eta\Lambda$ pairs is shown in Fig.~\ref{fig:models_eta}, where the ${J/\psi \Lambda}$ resonant structure has disappeared since the invariant $M_{J/\psi \Lambda}$ masses have been integrated out. The $\eta\Lambda$ invariant mass distributions have essentially the same shape as that of the distributions shown already in Fig.~6 of Ref.~\cite{feijoomagas}, which did not consider the additional contribution associated to the hidden charm strange pentaquark.  We see a broad peaked shape, associated to the NLO terms of the Lagrangian, which is more pronounced in the case of Model 1. As already discussed in Ref.~\cite{feijoomagas}, this structure is not associated to any resonant state since it appears at different energies in different channels and no pole in the complex plane was found either.

\begin{figure}[!htb]
\centering
  \includegraphics[width=\linewidth]{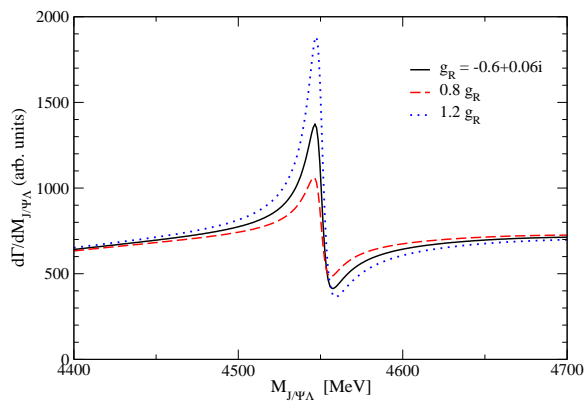}
\caption{(Color online) Invariant mass distributions of $J/\psi \Lambda$ states produced in the decay $\Lambda_b \to J/\psi ~ \eta \Lambda $, obtained for Model 2 and for different values of the coupling of the pentaquark to $J/\psi \Lambda$.}
  \label{fig:couplings_JPsi}
\end{figure}

\begin{figure}[!htb]
\centering
  \includegraphics[width=\linewidth]{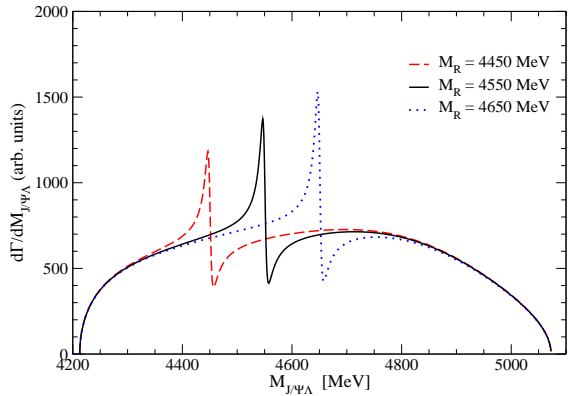}
\caption{(Color online) Invariant mass distributions of $J/\psi \Lambda$ states produced in the decay $\Lambda_b \to J/\psi ~ \eta \Lambda $, obtained for Model 2 and for different values of the pentaquark mass states.}
  \label{fig:mass_JPsi}
\end{figure}

In the following, results will be presented for only one model of the strong $\eta\Lambda$ interaction, chosen to be Model 2  as it provides a better account of the scattering observables  \cite{Feijoo:2015yja}.
The $J/\psi \Lambda$ invariant mass distributions displayed in Figs.~\ref{fig:couplings_JPsi} and \ref{fig:mass_JPsi},  for different values of the pentaquark coupling to $J/\psi\Lambda$ and for different values of the pentaquark mass, respectively, show obvious trends. From Fig.~\ref{fig:couplings_JPsi} we can conclude that the pentaquark could be seen over the background even if its coupling to $J/\psi \Lambda$  states were as low as $\mid g_{J/\psi \Lambda}\mid =0.48$. The unitary approaches of Refs.~\cite{Wu:2010jy,Wu:2010vk,Yang:2011wz,Xiao:2013yca} predict values for this coupling in between $0.5-1.0$, which 
make us believe that the strange pentaquark could leave a clear signature in the $J/\psi \Lambda$ mass spectrum.

The invariant mass distribution of $\eta\Lambda$ states is not sensitive to the characteristics of the pentaquark,  as already commented in the discussion of  Fig~\ref{fig:models_eta}. We have checked that changes in the coupling $\mid g_{J/\psi \Lambda}\mid$ or in the mass $M_R$ do not practically change the aspect of the $\eta\Lambda$ invariant mass spectrum.

\begin{figure}[!htb]
\centering
  \includegraphics[width=\linewidth]{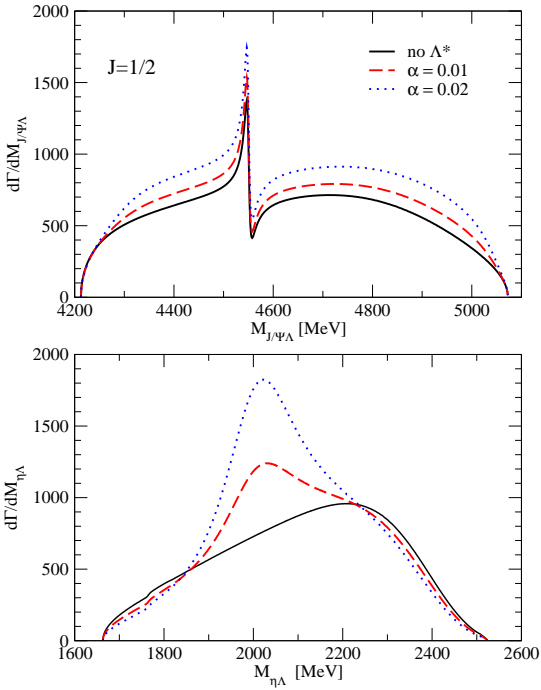}
\caption{(Color online) Invariant mass distributions of $J/\psi \Lambda$ (top panel) and $\eta\Lambda$ (bottom panel) states produced in the decay $\Lambda_b \to J/\psi ~ \eta \Lambda $, obtained from Model 2, assuming a pentaquark of $J^P=1/2^-$ and different strengths of the $\Lambda(2000)$ resonant contribution.}
  \label{fig:Lambdast}
\end{figure}

In Fig.~\ref{fig:Lambdast} we explore the effect of including the additional effect of a $\Lambda(2000)$ s-wave resonance coupling to $\eta\Lambda$ states. The unknown coupling strength $\alpha$ of Eq.~(\ref{eq:amplitude2}) is varied such that it produces a clearly visible change in the spectrum of  $\eta\Lambda$ invariant masses over what we obtain in the absence of this contribution, as seen in the bottom panel of Fig.~\ref{fig:Lambdast}. In the top panel we observe that the inclusion of the $\Lambda(2000)$ on the $J/\psi\Lambda$ pair distribution, where the $\eta\Lambda$ invariant masses have been integrated out,  essentially enhances the strength while keeping the same shape for the different values of $\alpha$.

\begin{figure}[!htb]
\centering
  \includegraphics[width=\linewidth]{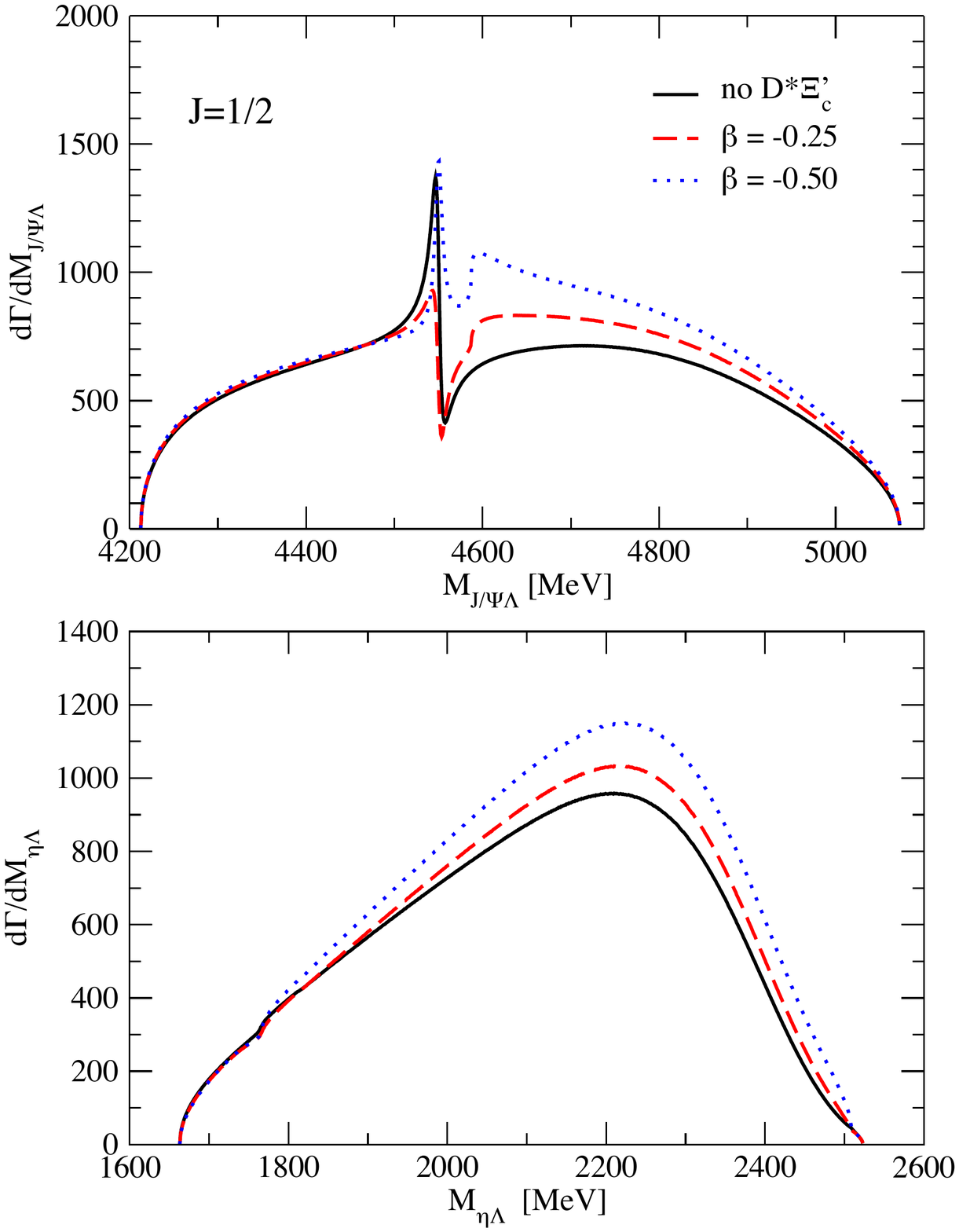}
\caption{(Color online) Invariant mass distributions of $J/\psi \Lambda$ (top panel) and $\eta\Lambda$ (bottom panel) states produced in the decay $\Lambda_b \to J/\psi ~ \eta \Lambda $, obtained from Model 2, assuming a pentaquark of $J^P=1/2^-$ and different strengths of the $\bar{D}^{*0}\Xi^\prime_c$ intermediate state contribution.}
  \label{fig:DstXiprime_m}
\end{figure}

\begin{figure}[!htb]
\centering
  \includegraphics[width=\linewidth]{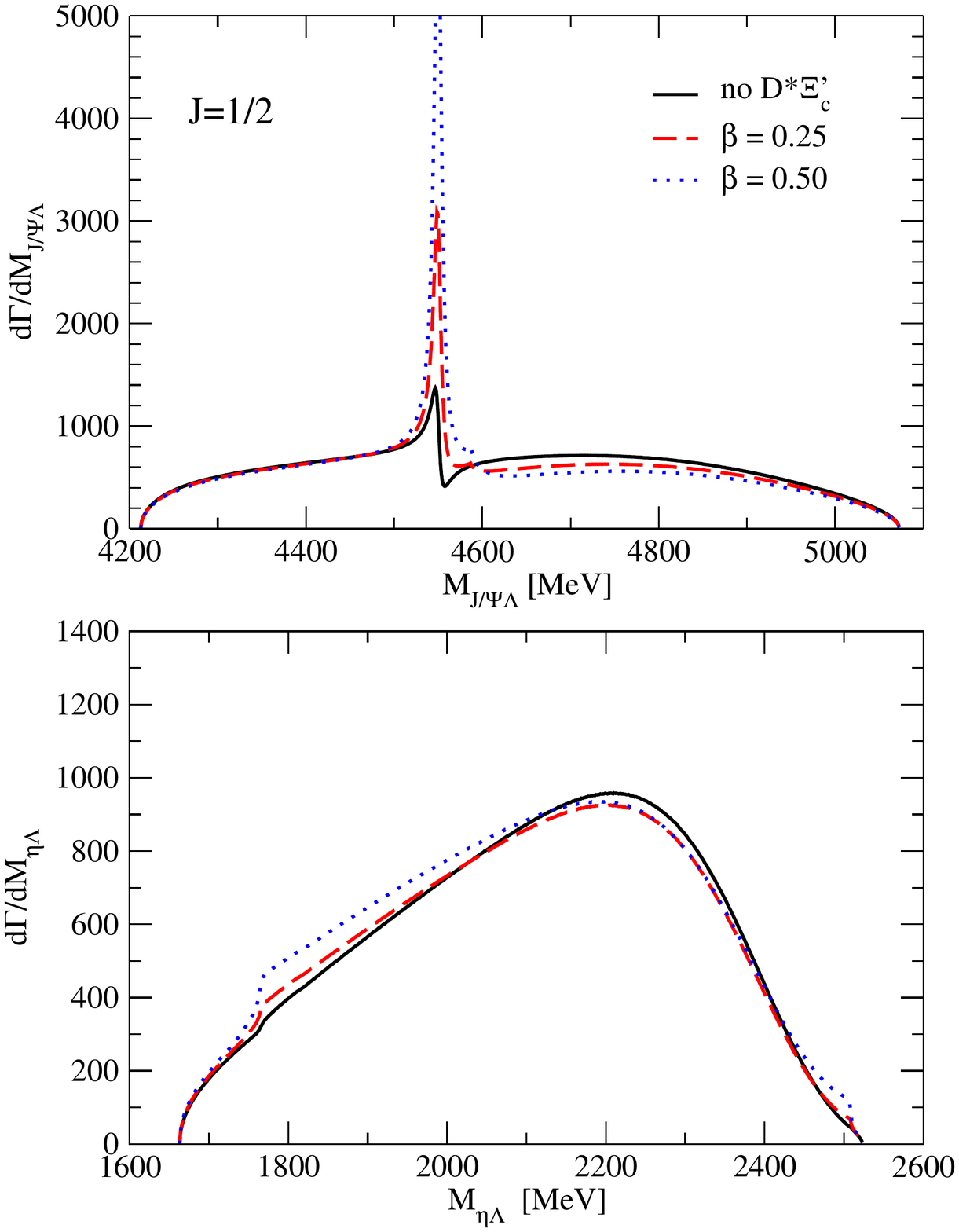}
\caption{(Color online) Invariant mass distributions of $J/\psi \Lambda$ (top panel) and $\eta\Lambda$ (bottom panel) states produced in the decay $\Lambda_b \to J/\psi ~ \eta \Lambda $, obtained from Model 2, assuming a pentaquark of $J^P=1/2^-$ and different strengths of the $\bar{D}^{*0}\Xi^\prime_c$ intermediate state contribution.}
  \label{fig:DstXiprime}
\end{figure}

We next explore the influence of the strange pentaquark being initiated by the excitation of a virtual $\bar D^{*0}\eta\Xi^\prime_c$ state, followed by the multiple scattering of $\bar D^{*0}\Xi^\prime_c$ leading to a final $J/\psi\Lambda$ pair and an $\eta$ meson. As discussed in Sect.~\ref{subsec:form}, the topology for this decay should lead to a reduced amplitude with respect to that of the $J/\psi\,\eta\Lambda$ case. We implement this phenomenologically through the parameter $\beta$, as seen in Eq.~(\ref{eq:amplitude2}), which is given the values $-0.5,-0.25,0.0, 0.25$, and $0.5$ accounting also for different relative sign cases. The results obtained with the negative values are displayed in Fig.~\ref{fig:DstXiprime_m} and those with the positive values in Fig.~\ref{fig:DstXiprime}. As seen in the top panel of Fig.~\ref{fig:DstXiprime_m}, the pentaquark signal for the negative values of $\beta$ gets somewhat reduced with respect to the case in which the virtual excitation of $\bar D^{*0}\eta\Xi^\prime_c$ states is omitted,  indicating a destructive interference with the direct excitation of $J/\psi\,\eta\Lambda$ states. However, the signal is still clearly visible over the background. The situation is completely different for the positive values of $\beta$. As seen on the top panel of  Fig.~\ref{fig:DstXiprime}, the signal of the pentaquark is tremendously enhanced due to a constructive interference between both mechanisms and to the fact that the coupling strength of the pentaquark to $\bar D^{*0}\Xi^\prime_c$ states is a factor 4 larger than that to 
$J/\psi\Lambda$. In both positive and negative $\beta$ cases, the changes seen in the $\eta\Lambda$ invariant mass spectra are relatively minor, as can be seen in the bottom panels of Figs.~\ref{fig:DstXiprime_m} and \ref{fig:DstXiprime}.

\begin{figure}[!htb]
\centering
  \includegraphics[width=\linewidth]{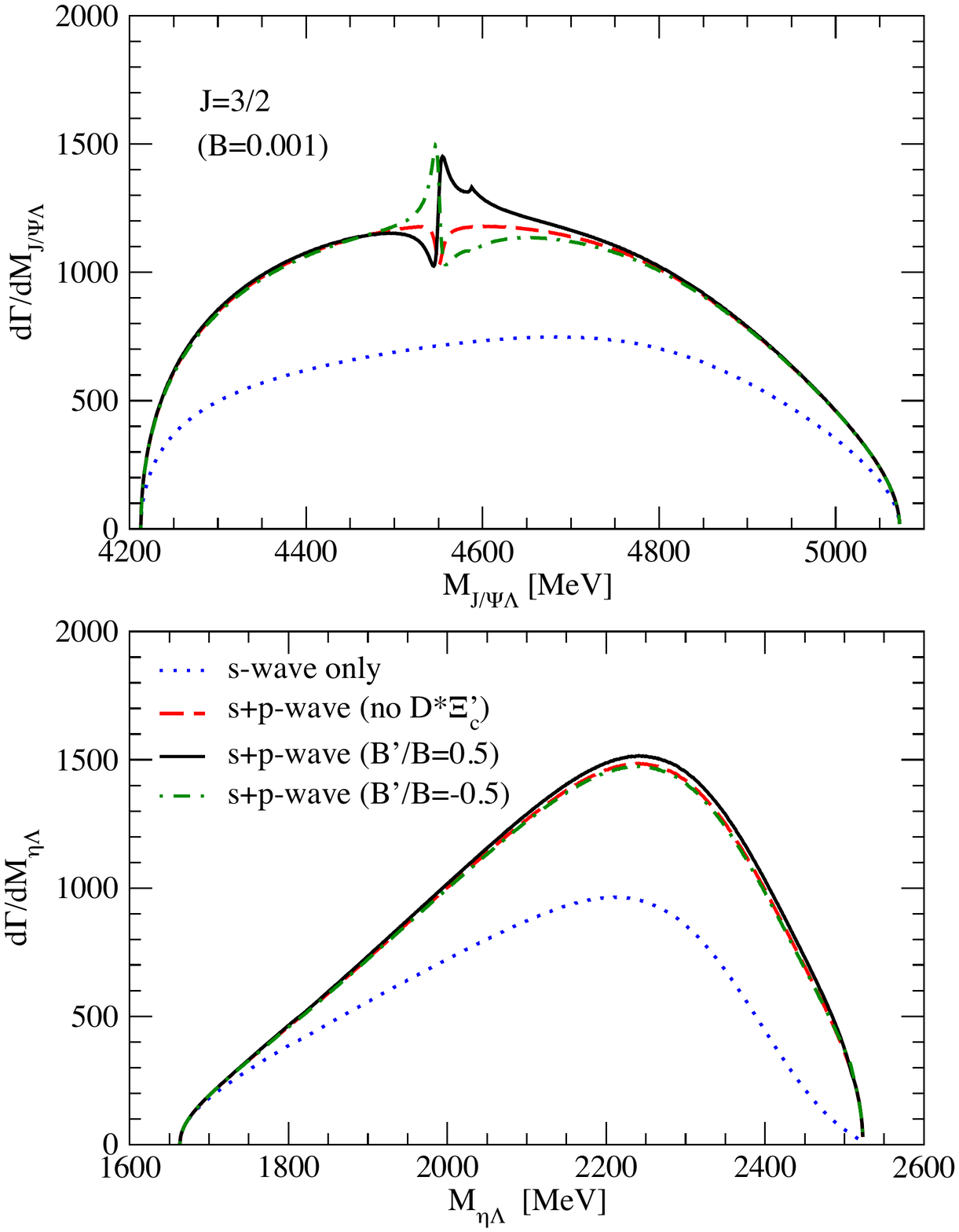}
    \setlength{\unitlength}{0.01\linewidth}
    \begin{picture}(100,0)    
       \put(50,102){\includegraphics[width=50\unitlength]{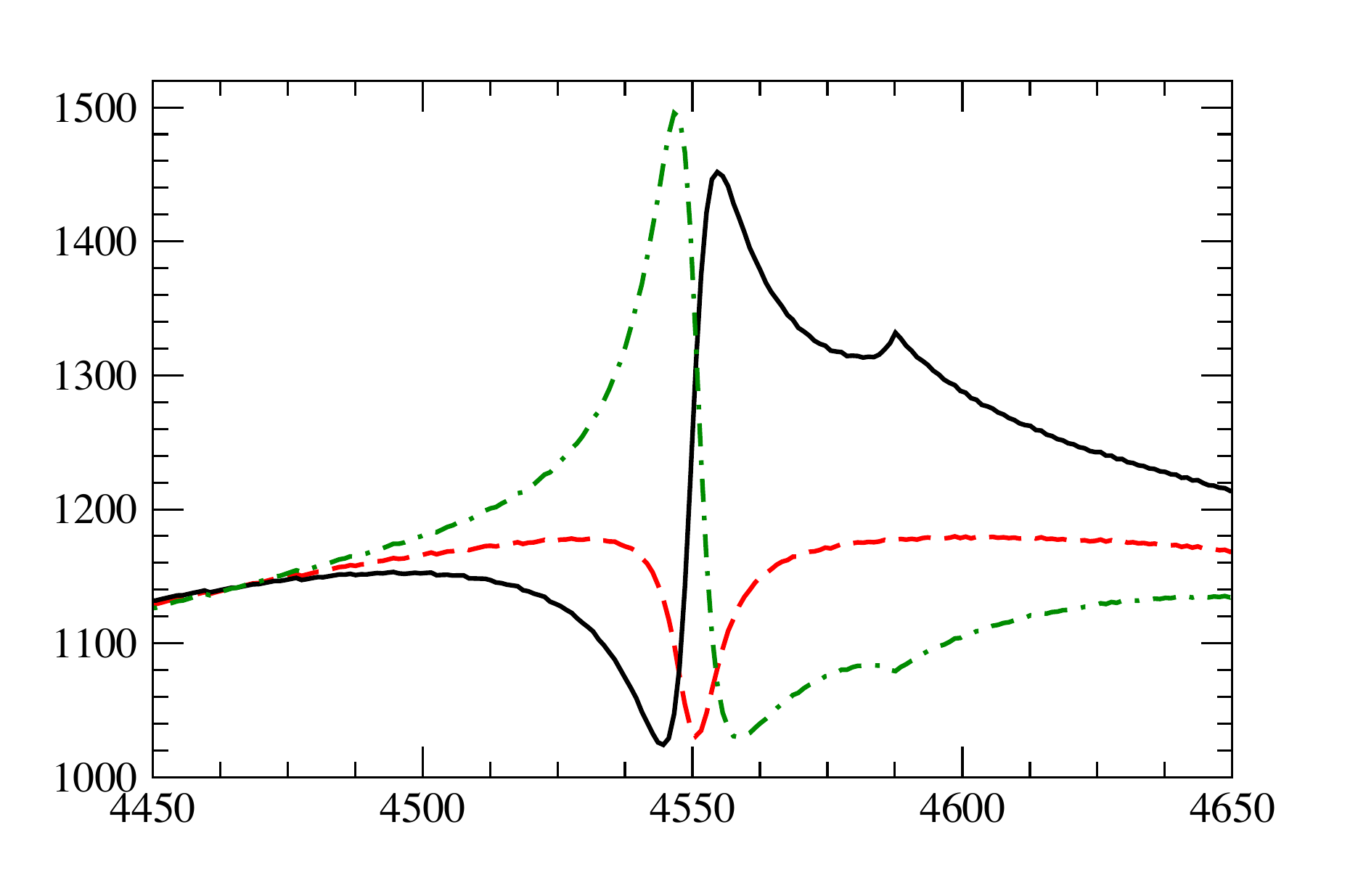}}
    \end{picture}
\caption{(Color online) Invariant mass distributions of $J/\psi \Lambda$ (top panel) and $\eta\Lambda$ (bottom panel) states produced in the decay $\Lambda_b \to J/\psi ~ \eta \Lambda $, obtained from Model 2, assuming a pentaquark of $J^P=3/2^-$. The dotted line is obtained with only an s-wave weak decay vertex, the dashed line also contains a p-wave contribution with $B=0.001$ MeV$^{-1}$, and the solid line implements the additional contribution of the $\bar{D}^{*0}\Xi^\prime_c$ intermediate state with $B^\prime/B=0.5$. The inset shows a zoomed view of the distribution in the region of the pentaquark mass. }
  \label{fig:caseJ32}
\end{figure}

Up to here, we have been discussing the results assuming the pentaquark to have $J^P=1/2^-$, which can then be produced by an s-wave mechanism for the $\Lambda_b$ decay. In the case of a $J^P=3/2^-$ pentaquark, which is another of the possibilities for the states predicted in \cite{Wu:2010jy,Wu:2010vk}, it is necessary to implement at least an additional  p-wave contribution, as that of Eq.~(\ref{eq:pwave}). Our results for this case are displayed in Fig.~\ref{fig:caseJ32}. 
The dotted line represents the case in which only the s-wave contribution is kept, producing a $J/\psi\Lambda$ pair in $1/2^-$. Since the pentaquark is now assumed to have $J^P=3/2^-$, it does not show in that $J/\psi \Lambda$ invariant mass spectrum, which reduces to a structureless background. 
We could have included, as in the study of the $J=1/2$ pentaquark case, the $\Lambda(2000)$ s-wave resonance contribution, but we have omitted this effect in the present $J=3/2$ discussion because, although it would be seen as an additional structure in the $\eta\Lambda$ invariant mass distribution,  it would simply contribute with a practically constant strength to the spectrum of $J/\psi \Lambda$ pairs, similarly to what we have found in Fig.~\ref{fig:Lambdast}.

When we add the p-wave vertex of Eq.~(\ref{eq:pwave}), we obtain the distributions displayed by the dashed curves in Fig.~\ref{fig:caseJ32}. The size of the coupling constant, $B=0.001$ MeV$^{-1}$, has been chosen so that the p-wave contribution has a visible effect over the s-wave $J/\psi \Lambda$ and $\eta\Lambda$ invariant mass distributions. The $J/\psi \Lambda$ spectrum, shown in  the top panel of Fig.~\ref{fig:caseJ32}, presents a dip at the pentaquak mass, which comes from the interference between the tree level and the $J/\psi \Lambda$ final state interaction terms, displayed by Figs.~\ref{fig:diagrams}(a) and (c), respectively, as can also be seen in Eq.~(\ref{eq:amp_pwave}). This is the same behavior as that observed in the study of the strange pentaquark from the $\Lambda_b\to J/\psi K^0 \Lambda$ decay in \cite{Lu:2016roh}. In the present work, we also incorporate the excitation of the pentaquark from the multiple scattering of  $\bar D^{*0}\Xi^\prime_c$ pairs produced in the virtual $\Lambda_b\to \bar D^{*0}\eta\Xi^\prime_c$ decay, which proceeds also in p-wave with a strength $B^\prime$. This is a necessary consideration if one wants to interpret the pentaquark as the state emerging from the interaction of $\bar D^{*0}\Xi^\prime_c$ and its related coupled states. If we now assume a ratio between the $\Lambda_b\to \bar D^{*0}\eta\Xi^\prime_c$ and  $\Lambda_b\to J/\psi\, \eta \Lambda$ amplitudes  of $B^\prime/B=0.5$, we obtain the solid curve, where the dip has turned into a wiggled shape. When the sign of $B^\prime$ is opposed to that of $B$, we find a similar behavior, although in a reflected way, as depicted by the dot-dashed curve. In either case, a visible pentaquark signal is obtained, as can be seen more clearly in the inset of Fig.~\ref{fig:caseJ32}.

\section{Conclusions}
\label{sec:conclusions}

   The recent finding of two structures in the $J/\psi p$ invariant mass distribution of the $\Lambda_b \to J/\psi K^- p$ decay, associated to two pentaquark states, together with its plausible explanation in terms of a previously predicted hidden charm baryon molecular state, prompted
us to study the decay of the $\Lambda_b$ into $ J/\psi \, \eta \Lambda$ final states. The $\Lambda_b \to  J/\psi \,\eta \Lambda$ decay, being a coupled channel of the $\Lambda_b \to J/\psi K^- p$ one, will occur with similar strength and one could observe, in the
$J/\psi \Lambda$ invariant mass spectrum, possible strange partners of the two non-strange pentaquark states reported by the LHCb collaboration.

We recall that when the hidden charm $N^*$ resonances were theoretically predicted as molecular states in several unitary approaches, some partner hidden charm strange $\Lambda^*$ states were also found. We have taken advantage of this finding and have predicted what signal should one of these states leave in the $\eta \Lambda$ and $J/\psi \Lambda$ invariant mass distributions of  the $\Lambda_b \to J/\psi \,\eta \Lambda$ reaction. We have found that, taking the values of the couplings of the hidden charm $\Lambda^*$ state to the $\bar D^{*0}\Xi^\prime_c$ and $J/\psi \Lambda$ channels obtained in the unitary approaches, one should observe clear and sizable peaks in the $J/\psi \Lambda$ mass distribution of the $\Lambda_b \to J/\psi\,\eta \Lambda$ decay. We have also used for this purpose the interaction of $\eta \Lambda$ with its coupled channels by means of a recent chiral unitary approach at next to leading order.

We have studied the dependence of our results on reasonable changes in the parameters of the models involved in our description of the process, as well as on the unknown properties of the speculated hidden charm strange pentaquark. We have observed that, while there appear changes in the position of the peak and in the shapes of the distributions, a resonance signal in the $J/\psi \Lambda$ invariant mass spectrum is clearly seen in all the cases. This gives us confidence that such an experiment should result into a successful proof of the existence of this new state and encourage the experimental analysis of this decay channel, for which our theoretical study predicts a similar strength than for the $\Lambda_b \to J/\psi K^- p$ reaction already analyzed by LHCb.

\begin{acknowledgements}
This work is partly supported by the Spanish Ministerio de Economia y Competitividad (MINECO) under the project MDM-2014-0369 of ICCUB (Unidad de Excelencia Mar\'\i a de Maeztu'), 
and, with additional European FEDER funds, under the contracts FIS2011-28853-C02-01, FIS2011-24154 and FIS2014-54762-P, by the Generalitat Valenciana in the program Prometeo II-2014/068,
by the Ge\-ne\-ra\-li\-tat de Catalunya contract 2014SGR-401,
and 
by the Spanish Excellence Network on Hadronic Physics FIS2014-57026-REDT.
\end{acknowledgements}


\begin{thebibliography}{}
%
%




\bibitem{Aaij:2015tga} 
  R.~Aaij {\it et al.} [LHCb Collaboration],
  Phys.\ Rev.\ Lett.\  {\bf 115}, 072001 (2015)
  [arXiv:1507.03414 [hep-ex]].

\bibitem{Aaij:2015fea} 
  R.~Aaij {\it et al.} [LHCb Collaboration],
  Chin.\ Phys.\ C {\bf 40}, no. 1, 011001 (2016)
  [arXiv:1509.00292 [hep-ex]].
  
\bibitem{Wu:2010jy} 
  J.~J.~Wu, R.~Molina, E.~Oset and B.~S.~Zou,
  Phys.\ Rev.\ Lett.\  {\bf 105}, 232001 (2010)
  [arXiv:1007.0573 [nucl-th]].
  
\bibitem{Wu:2010vk} 
  J.~J.~Wu, R.~Molina, E.~Oset and B.~S.~Zou,
  Phys.\ Rev.\ C {\bf 84}, 015202 (2011)
  [arXiv:1011.2399 [nucl-th]].
  
  
\bibitem{Yang:2011wz} 
  Z.~C.~Yang, Z.~F.~Sun, J.~He, X.~Liu and S.~L.~Zhu,
  Chin.\ Phys.\ C {\bf 36}, 6 (2012)
  [arXiv:1105.2901 [hep-ph]].
  
\bibitem{Xiao:2013yca} 
  C.~W.~Xiao, J.~Nieves and E.~Oset,
  Phys.\ Rev.\ D {\bf 88}, 056012 (2013)
  [arXiv:1304.5368 [hep-ph]].
  
\bibitem{Karliner:2015ina} 
  M.~Karliner and J.~L.~Rosner,
  Phys.\ Rev.\ Lett.\  {\bf 115}, no. 12, 122001 (2015)
  [arXiv:1506.06386 [hep-ph]].
  
\bibitem{Wang:2011rga} 
  W.~L.~Wang, F.~Huang, Z.~Y.~Zhang and B.~S.~Zou,
  Phys.\ Rev.\ C {\bf 84}, 015203 (2011)
  [arXiv:1101.0453 [nucl-th]].

  
\bibitem{Yuan:2012wz} 
  S.~G.~Yuan, K.~W.~Wei, J.~He, H.~S.~Xu and B.~S.~Zou,
  Eur.\ Phys.\ J.\ A {\bf 48}, 61 (2012)
  [arXiv:1201.0807 [nucl-th]].
  
\bibitem{Stone:2015iba} 
  S.~Stone,
  PoS EPS {\bf -HEP2015}, 434 (2015)
  [arXiv:1509.04051 [hep-ex]].

\bibitem{Chen:2015loa} 
  R.~Chen, X.~Liu, X.~Q.~Li and S.~L.~Zhu,
  Phys.\ Rev.\ Lett.\  {\bf 115}, no. 13, 132002 (2015)
  [arXiv:1507.03704 [hep-ph]].
 
\bibitem{Roca:2015dva} 
  L.~Roca, J.~Nieves and E.~Oset,
  Phys.\ Rev.\ D {\bf 92}, no. 9, 094003 (2015)
  [arXiv:1507.04249 [hep-ph]].
  
\bibitem{He:2015cea} 
  J.~He,
  Phys.\ Lett.\ B {\bf 753}, 547 (2016)
  [arXiv:1507.05200 [hep-ph]].
 
\bibitem{Meissner:2015mza} 
  U.~G.~Meissner and J.~A.~Oller,
  Phys.\ Lett.\ B {\bf 751}, 59 (2015)
  [arXiv:1507.07478 [hep-ph]].


\bibitem{Lebed:2015tna} 
  R.~F.~Lebed,
  Phys.\ Lett.\ B {\bf 749}, 454 (2015)
  [arXiv:1507.05867 [hep-ph]].
  
\bibitem{Maiani:2015vwa} 
  L.~Maiani, A.~D.~Polosa and V.~Riquer,
  Phys.\ Lett.\ B {\bf 749}, 289 (2015)
  [arXiv:1507.04980 [hep-ph]].
  
\bibitem{Anisovich:2015cia} 
  V.~V.~Anisovich, M.~A.~Matveev, J.~Nyiri, A.~V.~Sarantsev and A.~N.~Semenova,
  arXiv:1507.07652 [hep-ph].
 

\bibitem{Ghosh:2015ksa} 
  R.~Ghosh, A.~Bhattacharya and B.~Chakrabarti,
  arXiv:1508.00356 [hep-ph].
  
\bibitem{latest:diquark} 
  V.~V.~Anisovich, M.~A.~Matveev, J.~Nyiri, A.~V.~Sarantsev and A.~N.~Semenova,
  Int.\ J.\ Mod.\ Phys.\ A {\bf 30}, 1550190 (2015)
  [arXiv:1509.04898 [hep-ph]].
  
  
\bibitem{Chen:2015moa} 
  H.~X.~Chen, W.~Chen, X.~Liu, T.~G.~Steele and S.~L.~Zhu,
  Phys.\ Rev.\ Lett.\  {\bf 115}, no. 17, 172001 (2015)
  [arXiv:1507.03717 [hep-ph]].
  
\bibitem{Wang:2015epa} 
  Z.~G.~Wang,
  Eur.\ Phys.\ J.\ C {\bf 76}, no. 2, 70 (2016)
  [arXiv:1508.01468 [hep-ph]].
  
  
\bibitem{Scoccola:2015nia} 
  N.~N.~Scoccola, D.~O.~Riska and M.~Rho,
  Phys.\ Rev.\ D {\bf 92}, no. 5, 051501 (2015)
  [arXiv:1508.01172 [hep-ph]].
  
\bibitem{Guo:2015umn} 
  F.~K.~Guo, U.~G.~Meissner, W.~Wang and Z.~Yang,
  Phys.\ Rev.\ D {\bf 92}, no. 7, 071502 (2015)
  [arXiv:1507.04950 [hep-ph]].
   
\bibitem{Liu:2015fea} 
  X.~H.~Liu, Q.~Wang and Q.~Zhao,
  Phys.\ Lett.\ B {\bf 757}, 231 (2016)
  [arXiv:1507.05359 [hep-ph]].

\bibitem{Mikhasenko:2015vca} 
  M.~Mikhasenko,
  arXiv:1507.06552 [hep-ph].
  
\bibitem{Huang:2013mua} 
  Y.~Huang, J.~He, H.~F.~Zhang and X.~R.~Chen,
  J.\ Phys.\ G {\bf 41}, no. 11, 115004 (2014)
  [arXiv:1305.4434 [nucl-th]].
  
  
\bibitem{Garzon:2015zva} 
  E.~J.~Garzon and J.~J.~Xie,
  Phys.\ Rev.\ C {\bf 92}, no. 3, 035201 (2015)
  [arXiv:1506.06834 [hep-ph]].
  
  
\bibitem{Wang:2015jsa} 
  Q.~Wang, X.~H.~Liu and Q.~Zhao,
  Phys.\ Rev.\ D {\bf 92}, no. 3, 034022 (2015)
  [arXiv:1508.00339 [hep-ph]].
   
\bibitem{Kubarovsky:2015aaa} 
  V.~Kubarovsky and M.~B.~Voloshin,
  Phys.\ Rev.\ D {\bf 92}, no. 3, 031502 (2015)
  [arXiv:1508.00888 [hep-ph]].
 
\bibitem{Karliner:2015voa} 
  M.~Karliner and J.~L.~Rosner,
  Phys.\ Lett.\ B {\bf 752}, 329 (2016)
  [arXiv:1508.01496 [hep-ph]].
 
 
\bibitem{Cheng:2015cca} 
  H.~Y.~Cheng and C.~K.~Chua,
  Phys.\ Rev.\ D {\bf 92}, no. 9, 096009 (2015)
  [arXiv:1509.03708 [hep-ph]].
  
\bibitem{Li:2015gta} 
  G.~N.~Li, X.~G.~He and M.~He,
  JHEP {\bf 1512}, 128 (2015)
  [arXiv:1507.08252 [hep-ph]].
  
  
\bibitem{Mironov:2015ica} 
  A.~Mironov and A.~Morozov,
  JETP Lett.\  {\bf 102}, no. 5, 271 (2015)
  [arXiv:1507.04694 [hep-ph]].

\bibitem{Burns:2015dwa} 
  T.~J.~Burns,
  Eur.\ Phys.\ J.\ A {\bf 51}, no. 11, 152 (2015)
  [arXiv:1509.02460 [hep-ph]].

\bibitem{Chen:2016qju} 
  H.~X.~Chen, W.~Chen, X.~Liu and S.~L.~Zhu,
  arXiv:1601.02092 [hep-ph].
  
\bibitem{rocamai} 
  L.~Roca, M.~Mai, E.~Oset and U.~G.~Meissner,
  Eur.\ Phys.\ J.\ C {\bf 75}, no. 5, 218 (2015)
  [arXiv:1503.02936 [hep-ph]].

\bibitem{cascadeb} 
  H.~X.~Chen, L.~S.~Geng, W.~H.~Liang, E.~Oset, E.~Wang and J.~J.~Xie,
  arXiv:1510.01803 [hep-ph].

\bibitem{stone}
Private communication by Sheldon Stone

\bibitem{Lu:2016roh} 
  J.~X.~Lu, E.~Wang, J.~J.~Xie, L.~S.~Geng and E.~Oset,
   Phys.\ Rev.\ D {\bf 93}, no. 9, 094009 (2016)
  [arXiv:1601.00075 [hep-ph]].

\bibitem{Wang:2015pcn} 
  E.~Wang, H.~X.~Chen, L.~S.~Geng, D.~M.~Li and E.~Oset,
   Phys.\ Rev.\ D {\bf 93}, no. 9, 094001 (2016)
  [arXiv:1512.01959 [hep-ph]].

\bibitem{miyahara} 
  K.~Miyahara, T.~Hyodo and E.~Oset,
  Phys.\ Rev.\ C {\bf 92}, no. 5, 055204 (2015)
  [arXiv:1508.04882 [nucl-th]].

   
\bibitem{liang} 
  W.~H.~Liang and E.~Oset,
  Phys.\ Lett.\ B {\bf 737}, 70 (2014)
  [arXiv:1406.7228 [hep-ph]].
 
\bibitem{dai} 
  J.~J.~Xie, L.~R.~Dai and E.~Oset,
  Phys.\ Lett.\ B {\bf 742}, 363 (2015)
  [arXiv:1409.0401 [hep-ph]].


\bibitem{xie} 
  W.~H.~Liang, J.~J.~Xie and E.~Oset,
  Phys.\ Rev.\ D {\bf 92}, no. 3, 034008 (2015)
  [arXiv:1501.00088 [hep-ph]].

\bibitem{feijoomagas} 
  A.~Feijoo, V.~K.~Magas, A.~Ramos and E.~Oset,
  Phys.\ Rev.\ D {\bf 92}, no. 7, 076015 (2015)
  [arXiv:1507.04640 [hep-ph]].


\bibitem{bramon} 
  A.~Bramon, A.~Grau and G.~Pancheri,
  Phys.\ Lett.\ B {\bf 283}, 416 (1992).
  
\bibitem{Close:1979bt} 
  F.~E.~Close,
  ``An Introduction to Quarks and Partons'',
  Academic Press, London 1979, p 48
  
\bibitem{Feijoo:2015yja} 
  A.~Feijoo, V.~K.~Magas and A.~Ramos, 
 Phys.\ Rev.\ C {\bf 92}, 015206 (2015)
 [arXiv:1502.07956 [nucl-th]].
   
  \bibitem{PDG} 
  K.~A.~Olive {\it et al.}  [Particle Data Group Collaboration],
  Chin.\ Phys.\ C {\bf 38}, 090001 (2014).


\bibitem{Fernandez-Ramirez:2015tfa} 
  C.~Fernandez-Ramirez, I.~V.~Danilkin, D.~M.~Manley, V.~Mathieu and A.~P.~Szczepaniak,
  Phys.\ Rev.\ D {\bf 93}, no. 3, 034029 (2016)
  [arXiv:1510.07065 [hep-ph]].

\bibitem{Garzon:2012np} 
  E.~J.~Garzon and E.~Oset,
  Eur.\ Phys.\ J.\ A {\bf 48}, 5 (2012)
  [arXiv:1201.3756 [hep-ph]].

\bibitem{exp_data_scattering}
 J. K. Kim, Phys. Rev. Lett. {\bf14}, 89 (1965);
T.~S.~Mast, M.~Alston-Garnjost, R.~O.~Bangerter, 
     A.~S.~Barbaro-Galtieri, F.~T.~Solmitz and R.~D.~Tripp,
     Phys.\ Rev.\ D11 (1975) 3078;
 T. S. Mast, et al., Phys. Rev. D {\bf14}, 13 (1976);
  R. O. Bangerter, et al., Phys. Rev. D {\bf23}, 1484 (1981);
  J. Ciborowski, et al., J. Phys. G {\bf8}, 13 (1982);
  G. Burgun et al., Nucl. Phys. B {\bf8}, 447 (1968);
  J. R. Carlson, et al., Phys. Rev. D {\bf7}, 2533 (1973);
 P. M. Dauber, et al., Phys. Rev. {\bf179}, 1262 (1969);
  M. Haque et al., Phys. Rev. {\bf152}, 1148 (1966);
  G. W. London, et al., Phys. Rev. {\bf143}, 1034 (1966);
 T. G. Trippe, P. E. Schlein, Phys. Rev. {\bf158}, 1334 (1967);
 W. P. Trower, et al., Phys. Rev. {\bf170}, 1207 (1968);
    M.~Sakitt, T.~B.~Day, R.~G.~Glasser, N.~Seeman, J.~H.~Friedman, 
     W.~E.~Humphrey and R.~R.~Ross,
     Phys.\ Rev.\  139 (1965) B719;


\bibitem{exp_data_ratios}
 R. J. Nowak et al., Nucl. Phys. B {\bf 139}, 61 (1978);
D. N. Tovee et al., Nucl. Phys. B {\bf 33},  493 (1971).
  
\bibitem{Bazzi:2011zj} 
  M.~Bazzi, G.~Beer, L.~Bombelli, A.~M.~Bragadireanu, M.~Cargnelli, G.~Corradi, C.~Curceanu (Petrascu) and A.~d'Uffizi {\it et al.},
  Phys.\ Lett.\ B {\bf 704}, 113 (2011)
  [arXiv:1105.3090 [nucl-ex]].

\bibitem{ollerulf} 
  J.~A.~Oller and U.~G.~Meissner,
  Phys.\ Lett.\ B {\bf 500}, 263 (2001)
  [hep-ph/0011146].
 
\bibitem{Borasoy:2006sr} 
  B.~Borasoy, U.-G.~Meissner and R.~Nissler,
  Phys.\ Rev.\ C {\bf 74}, 055201 (2006)
  [hep-ph/0606108].


\bibitem{hyodonew} 
  Y.~Ikeda, T.~Hyodo and W.~Weise,
  Nucl.\ Phys.\ A {\bf 881}, 98 (2012)
  [arXiv:1201.6549 [nucl-th]].

\bibitem{Sharov:2011xq} 
  D.~A.~Sharov, V.~L.~Korotkikh and D.~E.~Lanskoy,
  Eur.\ Phys.\ J.\ A {\bf 47}, 109 (2011)
  [arXiv:1105.0764 [nucl-th]].

\bibitem{jackson} 
  B.~C.~Jackson, Y.~Oh, H.~Haberzettl and K.~Nakayama,
  Phys.\ Rev.\ C {\bf 91}, no. 6, 065208 (2015)
  doi:10.1103/PhysRevC.91.065208
  [arXiv:1503.00845 [nucl-th]].
 
\end{thebibliography}


\end{document}